\documentclass{lhep}       
\pdfoutput=1
\RequirePackage[T1]{fontenc}
\usepackage{anyfontsize}
\RequirePackage{fix-cm}
\usepackage{tikz}
\usepackage{tikz-cd}
\usetikzlibrary{arrows,decorations.markings}
\usepackage{genyoungtabtikz}
\usepackage{ytableau}
\usepackage{amssymb}
\usepackage{amsmath}
\usepackage{amsbsy}    
\RequirePackage{graphicx}
\RequirePackage{mathptmx}     
\RequirePackage{flushend}
\usepackage[titletoc,title]{appendix}
\usepackage[english]{babel}
\usepackage{pst-all}
\usepackage{longtable}
\usepackage{relsize}
\usepackage[numbers]{natbib}
\usepackage{amsfonts}
\usepackage[T1]{fontenc}
\usepackage[utf8]{inputenc}

\journal{LHEP}

\vol{xx}
\jyear{2021}
\pages{xxx} 

\received{xx }
\published{xx }

\def\bea{\begin{eqnarray}}
\def\eea{\end{eqnarray}}

\begin{document}

\title{Domain walls and M2-branes partition functions: M-theory and ABJM Theory 
}
\author{M.Nouman Muteeb }
\address{Abdus Salam School of Mathematical Sciences, Lahore, Pakistan}


\begin{abstract}
We study the BPS counting functions (free energies) of the M-string configurations.  We consider separated M5-branes along with  M2-branes stretched between them, with M5-branes acting as domain walls interpolating  different configurations of M2-branes. We find recursive structure in the free energies  of these configurations.
The M-string degrees of freedom on the domain walls are interpreted in terms of a pair of interacting supersymmetric WZW models. We also compute the  elliptic genus of the M-string in a toy model of the ABJM theory and compare it with the M-theory computation.
\end{abstract}
\maketitle
\tableofcontents
\newpage
\section{Domain walls in M-theory: an introduction}
The study of  classification and dynamics of the 6d Supersymmetric CFTs (SCFTs) is one of the important problems that is currently an active area of research. The maximally supersymmetric 6d CFTs are called $(2,0)$ theories. The type IIB string theory in the background of A-type
 gives rise to $(2,0)$ A-type theory. In the M-theory formulation the $(2,0)$ $A_{N-1}$ theory is the world volume theory of $N$ parallel and coincident M5-branes. Away from the conformal point the CFT describes the dynamics of the self-dual strings of  small tensions. In the M-theory these strings are described by the one dimensional intersections of M5-branes and M2-branes. The strings support $(4,0)$ quiver gauge theory. One crucial result of \cite{Haghighat:2013gba,Haghighat:2013tka,Hohenegger:2013ala,Hohenegger:2015cba,Hohenegger:2015btj} is 
that the elliptic genus of this quiver gauge theory turns out to be equal to the partition function of the bulk theory.\\
The superconformal group of the theory is $\mbox{Osp}(2,6|4)$. Let's denote  the 11d spacetime $\mathbb{R}^{1,10}$ by the coordinates $x^i,i=0,1,...,10$. The coincident $\mbox{M5}$-branes span the coordinates $\{x^0,x^1,...,x^5\}$. In the non-conformal limit the M5-branes are separated along the $x^6$ direction with positions denoted by $a_i,i=1,2,...,N$. The M2-brane are suspended between consecutive M5-branes and span the coordinates $\{x^0,x^1,x^6\}$. The M2-brane couples to a $2$-from field $B$ inside the M5-brane worldvolume and the boundary of the M2-brane is what is called the M-string. If we denote by $\Gamma^I$ the $32\times 32$ 11d gamma matrices, then the supersymmetries preserved by the M-string are given by
\bea
\Gamma^{016}\epsilon=\epsilon,\hspace{0.1cm} \Gamma^{012345}\epsilon=\epsilon,\hspace{0.1cm} \Gamma^{01}\epsilon=\epsilon.
\eea
where $\epsilon$ is the $32$-component spinor and $\Gamma^{I_1I_2...I_k}=\Gamma^{I_1}\Gamma^{I_2}...\Gamma^{I_k}$. \\
It is interesting to compactly $x^1$ to a circle of radius $R_1$ and consider M-theory compactification on this circle. This gives rise to $N=2^*$ SYM in the transverse five dimensions. The gauge coupling constant is given by $g_{YM}^2=4\pi^2R_1$. The momentum around the $S^1$ defines a quantum number of the 5d BPS particles $\frac{k}{R_1}=-\frac{1}{8g_{YM}^2}\int d^4x                       \mbox{tr}(F\wedge F)$ and with the corresponding mass $M=R_1\delta_{ij}+\frac{k}{R_1}$, where $\delta_{ij}$ denotes the separation between $i$-th and $j$-th M5-branes. Moreover the mass deformation also breaks the string worldsheet supersymmetry $(4,4)$ to $(4,0)$.\\
A further compactification of the theory can be considered along the compactified direction $x^0$. This makes the worldvolume of the M5-branes to be $\mathbb{R}^4\times \mbox{T}^2$. Twisting the theory as one moves around the second $S^1$ defines the so-called $\Omega$-background which makes it possible to apply the equivariant localisation to compute the partition function.
To engineer 5d $\mathcal{N}=1^*$ $SU(N)$ gauge theory one has to compactly M-theory on the elliptic CY3-fold $A_{N-1}\times \mbox{T}^2$. The instantons in the 4d theory are none other than the M-strings wrapped on the whole of $\mbox{T}^2$. \\

\begin{figure}
\begin{tikzpicture}[every node/.append style={midway}]

\draw[black, thick] (10,4) -- (10.6,4)node[anchor=west] {\quad $1$};
\draw[black, thick] (10,3) -- (9.3,3) node[anchor=east] {$2\quad$};
\draw[black, thick] (10,4) -- (10,3)node[anchor=east] {$Q$};

\draw[black, thick] (10,4) -- (9.5,4.5);

\draw[black, thick] (9.5,4.5) -- (9.5,5.3);
\draw[black, thick] (9.5,4.5) -- (8.8,4.5)node[anchor=east] {$1\quad$};
\draw[black, thick] (9.5,4.5) -- (8.8,4.5)node[anchor=south] {$Q_{\tau}\quad$};

\draw[black, thick] (10,3) -- (10.5,2.5);

\draw[black, thick] (10.5,2.5) -- (11.2,2.5)node[anchor=west] {\quad $2$};
\draw[black, dotted] (10.5,2.5) -- (10.5,1.7);
\draw[black, dotted] (10.5,2.5) -- (10.5,1.7);

\draw[black, thick] (10.5,1.7) -- (11.1,1)node[anchor=west] {$Q_m$};
\draw[black, thick] (10.5,1.7)-- (9.8,1.7)node[anchor=east] {$N\quad$};
\draw[black, thick] (11.1,1)-- (11.8,1)node[anchor=west] {\quad $N$};
\draw[black, thick] (11.1,1)-- (11.1,0.3);

\end{tikzpicture}

 \caption{horizontally compactified toric web with corresponding fugacities $Q_{\tau}=e^{2\pi i \tau},Q_{m}=e^{2\pi i m}$ and $Q$.    }
    \label{N2}
\end{figure}
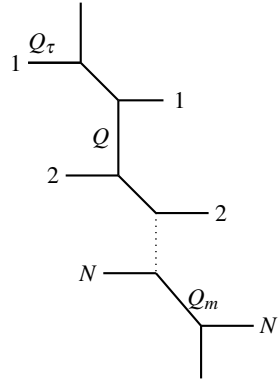

 Recall that the symmetry  $U(1)_{\epsilon_1}\times U(2)_{\epsilon_2}\times U(1)_{m}$ acts on the two $\mathbb{R}^4$s defined by the coordinates $\mathbb{R}^4_{||}:z_1=x_2+ix_3,z_2=x_4+ix_5$ and $\mathbb{R}^4_{\perp}:z_3=x_7+ix_8,z_4=x_9+ix_{10}$ as follows:
$(z_1,z_2)\to (q z_1,t^{-1}z_2),
(z_3,z_4)\to (\sqrt{\frac{t}{q}}e^{\pi i (2m)}z_3,\sqrt{\frac{t}{q}}e^{\pi i(-2m)}z_4)$
where $q:=e^{2\pi i \epsilon_1},t:=e^{-2\pi i \epsilon_2}$.
The toric geometry, dual to the type IIB D5-NS5-(1,1) branes web, underlying the M-string computation is given in figure \ref{N2}. The diagonal edges  correspond to the mass parameter $m$,the horizontal direction is periodic with period $\tau$ and $Q$ is the fugacity corresponding to the internal vertical lines. 

\begin{figure}
\begin{tabular}{ |p{3cm}|p{4cm}|  }
 \hline
 \multicolumn{2}{|c|}{11d M-theory space-time} \\
 \hline
 & $x^0\hspace{0.1cm} x^1 \hspace{0.1cm} x^2\hspace{0.1cm} x^3\hspace{0.1cm} x^4\hspace{0.1cm} x^5\hspace{0.1cm} x^6\hspace{0.1cm} x^7\hspace{0.1cm} x^8\hspace{0.1cm} x^9\hspace{0.1cm} x^{10}$\\
 \hline
 M5   & $\times\hspace{0.1cm} \times \hspace{0.1cm} \times\hspace{0.1cm} \times\hspace{0.1cm} \times\hspace{0.1cm} \times\hspace{0.1cm} \hspace{0.1cm} \hspace{0.1cm} \hspace{0.1cm} \hspace{0.1cm} $ \\
 M2&   $\times\hspace{0.1cm} \times \hspace{0.1cm} \hspace{0.1cm} \hspace{0.1cm} \hspace{0.1cm} \hspace{0.1cm} \hspace{0.1cm} \hspace{0.1cm} \qquad\hspace{0.18cm}\times \hspace{0.1cm} $ \\
 M-string &$\times\hspace{0.1cm} \times \hspace{0.1cm} \hspace{0.1cm} \hspace{0.1cm} \hspace{0.1cm} \hspace{0.1cm} \hspace{0.1cm} \hspace{0.1cm} \hspace{0.1cm} \hspace{0.1cm} $ \\
$ \epsilon_1$ &$\hspace{0.1cm}\hspace{0.1cm} \hspace{0.1cm} \hspace{0.1cm} \times \hspace{0.1cm}\times \hspace{0.1cm} \hspace{0.1cm} \hspace{0.1cm} \hspace{0.1cm}\hspace{0.1cm}\hspace{0.1cm}\hspace{0.1cm}\times\hspace{0.1cm}\times\hspace{0.1cm}\times\hspace{0.1cm}\times $ \\
$ \epsilon_2$ &$\hspace{0.1cm}\hspace{0.1cm} \hspace{0.1cm} \hspace{0.1cm}  \hspace{0.1cm} \hspace{0.1cm} \hspace{0.1cm}\hspace{0.1cm}\times \hspace{0.1cm}\times\hspace{0.1cm} \hspace{0.1cm}\hspace{0.1cm}\times\hspace{0.1cm}\times\hspace{0.1cm}\times\hspace{0.1cm}\times $ \\
$ m$ &$\hspace{0.1cm}\hspace{0.1cm} \hspace{0.1cm} \hspace{0.1cm}\hspace{0.1cm}\hspace{0.1cm}  \hspace{0.1cm} \hspace{0.1cm} \hspace{0.1cm}\hspace{0.1cm} \hspace{0.1cm}\hspace{0.1cm} \hspace{0.1cm}\hspace{0.1cm}\times\hspace{0.1cm}\times\hspace{0.1cm}\times\hspace{0.1cm}\times $ \\
 \hline
\end{tabular}
\caption{M-theory vacuum}
 \label{MTV}
\end{figure}

The target space of the gauged linear sigma model on the worldsheet of M-string is the singular space $\mbox{Sym}^{n}\mathbb{R}^4/S^n$ described as the configuration space of $n$ points on $\mathbb{R}^4$ moded by the permutation group $S_n$.The singularity is due to the coincidence of multiple points and resolving this singularity gives rise to the Hilbert scheme of $n$ points on $\mathbb{C}^2$. So   instead of dealing with the ill defined sigma model on  $\mbox{Sym}^{n}\mathbb{R}^4/S^n$, one can work with a $(4,0)$ sigma model on the $\mathcal{M}=\mbox{Hilb}^n(\mathbb{C}^2)$ \cite{Haghighat:2013gba}. The matter content of the theory is given by the sections of different bundles on the instanton moduli space. In the current situation we have tangent bundle $T_{\mathcal{M}}$ and the complex tautological bundle $E$. The bundle $E$ corresponds to the contribution of fundamental hypermultiplet in the gauge theory instanton computation.  In this theory the left handed fermions are the sections of tangent bundle $T_\mathcal{M}$, whereas the right handed fermions are the sections of $E\oplus E^{*}$. The combination $E\oplus E^{*}$, corresponding to the two hypermultiplets, signifies the fact that in the  toric geometry the $\mathbb{P}^1$ of the second hypermultplet is flopped \cite{Haghighat:2013gba}.
\\

The M-string worldsheet theory  defines  the target space $\mbox{Hilb}^n(\mathbb{C}^2)$.

The coupling of the left moving fermions and right moving fermions to different bundles correspond to the different boundary conditions  around the 1-cycles of $\mbox{T}^2$ when computing the elliptic genus. The $(2,2)$ sigma model contains bosons $\phi^i$ and the  fermions $\psi^i_-,\psi^{\bar{i}}_-,\psi^i_+\psi^{\bar{i}}_+$. Locally the bosons describe a map $\Phi:world-sheet\to\mathcal{M}$. If we denote by $K$ the bundle of $(1,0)$-forms and by $\bar{K}$ the bundle of $(0,1)$ forms then the fermions are defined by the following pullback maps
\bea
\psi^i_-&=&K^{\frac{1}{2}}\otimes\Phi^*T_{\mathcal{M}},\quad \psi^{\bar{i}}_-=K^{\frac{1}{2}}\otimes\Phi^*\bar{T}_{\mathcal{M}}\nonumber\\
\psi^i_+&=&\bar{K}^{\frac{1}{2}}\otimes\Phi^*T_{\mathcal{M}},\quad \psi^{\bar{i}}_+=\bar{K}^{\frac{1}{2}}\otimes\Phi^*\bar{T}_{\mathcal{M}}
\eea
If right moving fermions $\eta_a$ are also included the supersymmetry gets broken to $(2,0)$ and we define
\bea
\eta^a=\bar{K}^{\frac{1}{2}}\otimes\Phi^*(E\oplus E^*)
\eea
\\
In mathematical terms the elliptic genus is defined by the index of the following formal combination of vector bundles $V_{Q_{\tau},y}$ \cite{Haghighat:2013gba,Hohenegger:2015cba}
\bea
&V_{Q_{\tau},Q_m}=\bigotimes_{k=0}^{\infty}\wedge_{-Q_mQ_{\tau}^{k-1}}(E\oplus E^*)\otimes \bigotimes_{k=1}^{\infty}\wedge_{-Q_m^{-1}Q_{\tau}^{k}}(E\oplus E^*)^{*}\otimes\nonumber\\& \bigotimes_{k=1}^{\infty}S_{Q_{\tau}^k}T^*_{\mathcal{M}}\otimes  \bigotimes_{k=1}^{\infty}S_{Q_{\tau}^k}T_{\mathcal{M}}\nonumber\\
\eea
where  $\wedge_x W=\sum_{k\ge 0}(\wedge^k W)x^k$ and $S_xW=\sum_{k\ge 0}(S^k W)x^k$  are formal power series. They define the exterior powers  and the symmetric powers of a bundle $W$ as coefficients.
The partition function  $Z$ of the M-string configuration is the generating function of the elliptic genus $\chi(M,V_{Q_{\tau},y})$

\bea
&Z=\sum_{k\ge 0}Q^{k}\chi(M,V_{Q_{\tau},y},k)=\sum_{k\ge 0}Q^{k}\int_{\mathcal{M}}\mbox{ch}((E\oplus E^*)_{Q_{\tau},y})\mbox{Td}(T_{\mathcal{M}})\nonumber\\&=\sum_{k\ge 0}(Q_mQ)^{k}\int_{\mathcal{M}}e^{\sum_{i=1}^{d}\frac{1}{2}(x_i-\tilde{x}_i)}\prod_{i=1}^d\frac{\theta_1(\tau;-m+\frac{\tilde{x_i}}{2\pi i})}{\theta_1(\tau;\frac{x_i}{2\pi i})}\nonumber\\
\eea
where $\mathcal{M}=\mbox{Hilb}^n[\mathbb{C}^2]$,$ \tilde{x}_i,x_i$ denote the formal roots of the Chern polynomial of the bundles $E\oplus E^*,T\mathcal{M}$ respectively, $\mbox{ch}(V)$ is the Chern character of the bundle $V$, $\mbox{Td}(T)$ is the Todd class of the tangent bundle,  and in the second equality of the last expression we have used the identity 
\bea
\mbox{ch}(\mbox{V}_{Q_{\tau},Q_m})\mbox{Td}(\mbox{T}_{\mathcal{M}})=\prod_{k=1}^{\infty}\frac{\prod_{i=1^r}(1-Q_{\tau}^{k-1}Q_me^{-\tilde{x}_i})(1-Q_{\tau}^{k}Q_m^{-1}e^{\tilde{x}_i})}{\prod_{j=1}^dx_j^{-1}(1-Q_{\tau}^{k-1}e^{-x_j})(1-Q_{\tau}^{k}e^{x_j})}
\eea
 along with the following definition of Dedekind theta function $\theta_1(\tau)$ 
\bea
&\theta_1(\tau;z)=-ie^{\frac{i\pi \tau}{4}}e^{i\pi z}\prod_{k=1}^{\infty}(1-e^{2\pi i k\tau})(1-e^{2\pi i k\tau}e^{2\pi i z})\nonumber\\&(1-e^{2\pi i (k-1)\tau}e^{-2\pi i z})\nonumber\\
&\theta_1(\tau+1;z)=\theta_1(\tau;z),\hspace{0.1cm} \theta_1(-\frac{1}{\tau};\frac{z}{\tau})=-i(i\tau)^{\frac{1}{2}}e^{\frac{i\pi z^2}{\tau}}\theta_1(\tau;z)
\eea
\\
Recall that $V=E\oplus E^*$ is  a bundle on $\mbox{Hilb}^k[\mathbb{C}^2]$. If an ideal $\mbox{I}$ denotes a point of $\mbox{Hilb}^k[\mathbb{C}^2]$ then the fiber of the bundle $V$ over $\mbox{I}$ was found to be \cite{Haghighat:2013gba}
\bea
V|_I=\mbox{Ext}^1(\mathcal{O},\mbox{I})\otimes L^{-\frac{1}{2}}\oplus \mbox{Ext}^1(\mbox{I},\mathcal{O})\otimes L^{-\frac{1}{2}}
\eea
where $L$ is the canonical line bundle on $\mathbb{C}^2$. It has been shown that the appearance of the Ext-groups  is related \cite{Katz:2002gh} to the counting of  open string states between the D-branes wrapped on the holomorphic submanifolds. By determining the equivariant weights of the bundle $V$ the M-string partition function is determined. Intuitively each $M5$-brane with M2-branes ending on the left and right  gives rise to a factor $\mbox{Ext}^1(\mbox{I},\mbox{J})\otimes L^{-\frac{1}{2}}$ in the bundle with $\mbox{I}$ denoting a point of $\mbox{Hilb}^n[\mathbb{C}^2]$ corresponding to the M2-brane on the left and $\mbox{J}$ denotes a point of $\mbox{Hilb}^m[\mathbb{C}^2]$ corresponding to the M2-brane on the right.\\
For  general moduli space $\mathcal{M}_{k_1,...,k_{N-1}}=\mbox{Hilb}^{k_1}[\mathbb{C}^2]\times \mbox{Hilb}^{k_2}[\mathbb{C}^2]\times...\times \mbox{Hilb}^{k_{N-1}}[\mathbb{C}^2]$
the fiber of the corresponding bundle $V$ over $(\mbox{I}_1,...,\mbox{I}_{N-1})\in \mathcal{M}_{k_1,...,k_{N-1}}$ is given by
\bea
V|_{(\mbox{I}_1,...,\mbox{I}_{N-1})}=\big( \oplus_{a=0}^{N-1}\mbox{Ext}^1(\mbox{I}_a,\mbox{I}_{a+1})\otimes L^{-\frac{1}{2}} \big)
\eea
The fixed points are in one-to-one correspondence with the set of partitions $(\nu_1,...,\nu_{N-1})$ and the equivariant weights of $V$ over the fixed point are
\bea
&\{Q_mq^{-i+\frac{1}{2}}t^{j-\frac{1}{2}}|(i,j)\in\nu_1 \}\cup \{Q_mq^{i-\frac{1}{2}}t^{-j+\frac{1}{2}}|(i,j)\in\nu_{N-1} \}\nonumber\\
&\bigg(\cup_{a=1}^{N-2}\{Q_mq^{\nu^t_{a,j}-i+\frac{1}{2}}t^{\nu_{a+1,i}-j+\frac{1}{2}}|(i,j)\in\nu_a\} \cup\nonumber\\&\{Q_mq^{-\nu^t_{a+1,j}+i-\frac{1}{2}}t^{-\nu_{a,i}+j-\frac{1}{2}}|(i,j)\in\nu_{a+1}\}  \bigg)\nonumber\\
\eea
Using these weights at the fixed points the partition function turns out to be

\bea\label{ZNTOP}
&Z_{N}(\tau,m,t_{f_1},t_{f_2},...,t_{f_N},\epsilon_1,\epsilon_2)
\nonumber\\
&=\sum_{k_1,k_2,...,k_{N-1}}(\prod_{a=1}^{N-1}(-Q_{f_a})^{|\nu_a|})\sum_{|\nu_1|=k_1,...,|\nu_{N-1}|=k_{N-1}}\nonumber\\&\times\prod_{a=1}^{N-1}\prod_{(i,j)\in \nu_a}\frac{\theta_1(\tau;z_{i,j}^a)\theta_1(\tau;v_{i,j}^a)}{\theta_1(\tau;w_{i,j}^a)\theta_1(\tau;u_{i,j}^a)}\nonumber\\
\eea
where
\bea
z^a_{i,j}&=&-m+\epsilon_1(-\nu^t_{a-1,j}+i-\frac{1}{2})-\epsilon_2(-\nu_{a,i}+j-\frac{1}{2}),\nonumber\\
 v^a_{i,j}&=&-m+\epsilon_1(\nu^t_{a+1,j}-i+\frac{1}{2})-\epsilon_2(\nu_{a,i}-j+\frac{1}{2}),\nonumber\\
w^a_{i,j}&=&\epsilon_1(\nu^t_{a,j}-i)-\epsilon_2(\nu_{a,i}-j+1),\nonumber\\
 u^a_{i,j}&=&\epsilon_1(-\nu^t_{a,j}+i-1)-\epsilon_2(-\nu_{a,i}+j)\nonumber\\
 \nu_{0}&=&0,\hspace{0.1cm}  \nu_{N}=0.
\eea
and $Q_{f_a}:=Q_a=e^{2\pi i t_{f_a}},a=1,...,N-1$ are the fugacities in terms of the coulomb branch parameters $t_{f_a}$ that determine the distance between consecutive M5-branes. From the expression  (\ref{ZNTOP})  we can isolate the following part
\bea
Z^{}_{k_1,k_2,...,k_{N-1}}=\sum_{|\nu_1|=k_1,...,|\nu_{N-1}|=k_{N-1}}\prod_{a=1}^{N-1}\prod_{(i,j)\in \nu_a}\frac{\theta_1(\tau;z_{i,j}^a)\theta_1(\tau;v_{i,j}^a)}{\theta_1(\tau;w_{i,j}^a)\theta_1(\tau;u_{i,j}^a)}\nonumber\\
\eea
which can be interpreted \cite{Gomis:2008vc} as the partition function of the following configuration  of the wrapped M2-branes: $k_1$ M2-branes between 1st and 2nd M5-branes, $k_2$ M2-branes between 2nd and 3rd M5-branes and so on upto $k_{N-1}$ M2-branes between $(N-2)$-th and $(N-1)$-th M5-branes. \\

After stripping off the gauge theory $U(1)$ part $N\mbox{PLog}Z_1$ from the free energy
\bea
\Omega_N(\tau,m,t_{f_a},\epsilon_1,\epsilon_2)&=&\mbox{PLog}Z_N(\tau,m,t_{f_a},\epsilon_1,\epsilon_2)\nonumber\\&=&\mbox{NPLog}Z_1+\mbox{PLog} \tilde{Z}_N
\eea
one can expand the  remaining free energy
   $\tilde{\Omega}(\tau,m,t_{f_a},\epsilon_1,\epsilon_2):=\mbox{PLog}\tilde{Z}_{N}$  in terms of the fugacities $Q_{f_a}$ as follows
\bea
\tilde{\Omega}(\tau,m,t_{f_a},\epsilon_1,\epsilon_2):&=&\mbox{PLog}\tilde{Z}_{N}\nonumber\\&=&\sum_{k_i=1}Q_{f_1}^{k_1}...Q_{f_{N-1}}^{k_{N-1}}F_{k_1,k_2,...,k_{N-1}}(\tau,m,\epsilon_1,\epsilon_2)\nonumber\\
\eea
where the multi-index function  $F_{k_1,k_2,...,k_{N-1}}(\tau,m,\epsilon_1,\epsilon_2)$ counts the degeneracies of the M-strings bound states. 

\section*{Presentation of the article}
 After briefly introducing the M-strings and the corresponding elliptic genus of its worksheet theory in section \ref{recursivestructure}, we discuss recursive structure in the expressions for free energies corresponding to various configuration of the M2-M5 branes. A general configuration consists of an array of multiple M2-branes sandwitched between M5-branes. The M2-branes vacua are labelled by the tuple of integer partitions that correspond to the Young diagrams transforming in different representations. We discuss M2-M5 branes configurations in which the M2-branes  are labelled by antisymmetric representations and symmetric representations. For these representations the free energies enjoy a partial  recursive structure. For mixed\footnote{configurations of M5-M2 branes in which an M2-brane may carry a symmetric or an antisymmetric representation.} representations the recursive structure is lost except for the configuration shown in section 3.
   In section \ref{wzw} it is discussed that the open topological string wave function for the configuration  M2-M5-M2 of branes can be described in terms of  two WZW models coupled together. In section  \ref{MABJM} we compute the elliptic genus for the M-strings that arise in ABJM model. We compare it to the M-string elliptic genus as computed in M-theory framework.
\section{Recursive structure in the M-strings partition function}\label{recursivestructure}

For the M-string configuration in which a single M2-brane is stretched between consecutive M5-branes, the free energies show interesting recursive structure \cite{Hohenegger:2015cba}. For more complicated configurations the recursive structure is not apparent in the  expression for  free energies and 
the correct objects to decompose are the components  $Z_{\nu_1\nu_2...\nu_n}$.In doing the following computations we will often use the following symmetry of the elliptic genera indices $\{\mu_1,...,\mu_n,\emptyset_1,\emptyset_2,...,\emptyset_m\}$
\bea
Z_{permutation\{\hspace{0.1cm}\mu_1,...,\mu_n,\emptyset_1,\emptyset_2,...,\emptyset_m \}}=Z_{\mu_1,...,\mu_n,\emptyset_1,\emptyset_2,...,\emptyset_m }
\eea
where $\emptyset_i=\emptyset_j$ for all $i,j$, $m$ can be less than, equal to or greater than $n$ and $permutation$ denotes any possible permutation of the given indices.\\ For the configuration of partitions $\{\alpha,\alpha,\alpha,...,\alpha\}$
\bea
Z_{\alpha\alpha..\alpha}= all\hspace{0.1cm}  possible\hspace{0.1cm} ways \hspace{0.1cm} of\hspace{0.1cm} factorizing\hspace{0.1cm}+Z_{\alpha} W_{\alpha}^{(k-1)}
\eea
where $W_{\alpha}$ can be thought of as a universal factor corresponding to removing a single M5-brane.
The meaning of the phrase $``all\hspace{0.1cm}  possible\hspace{0.1cm} ways \hspace{0.1cm} of\hspace{0.1cm} factorizing "$ is the following: we will see explicitly in the next section that  there is a recursive structure in the expansion coefficients $Z_{\nu_1\nu_2...\nu_n}$ of the elliptic genera ; for example
\bea
Z_{222}-2Z_{22\emptyset}Z_{2\emptyset\emptyset}+Z_{2\emptyset\emptyset}^3=Z_{2\emptyset\emptyset}W_2(\tau,m,\epsilon_1,\epsilon_2)
\eea
where the explicit expressions for the factors  $Z_{222},Z_{22\emptyset},Z_{2\emptyset\emptyset}$ and $W_2(\tau,m,\epsilon_1,\epsilon_2)$ are given in the next section.
We can also write the last expression  as
\bea\label{eq:allfactors}
Z_{222}&=&2Z_{22\emptyset}Z_{2\emptyset\emptyset}-Z_{2\emptyset\emptyset}^3+Z_{2\emptyset\emptyset}W_2(\tau,m,\epsilon_1,\epsilon_2)\nonumber\\
&=& all\hspace{0.1cm}  possible\hspace{0.1cm} ways \hspace{0.1cm} of\hspace{0.1cm} factorizing\hspace{0.1cm}\nonumber\\&+&Z_{2\emptyset\emptyset}W_2(\tau,m,\epsilon_1,\epsilon_2)
\eea

In general, if there are $k$ partitions $\alpha_1,...,\alpha_k$ of the same size then it is the case that:
\bea\label{eq:fusion}
Z_{\alpha_1\alpha_2...\alpha_k}&=& all\hspace{0.1cm} possible\hspace{0.1cm} ways\hspace{0.1cm} of\hspace{0.1cm} factorizing\nonumber\\&+& Z_{\beta} W_{\alpha_1\alpha_2} W_{\alpha_2\alpha_3}....W_{\alpha_{k-2}\alpha_{k-1}}W_{\alpha_{k-1}\alpha_{k}}
\eea
where $\beta$ is the result of  fusing \cite{Gaiotto:2008ak}  the partitions $(\alpha_1,\alpha_2,...,\alpha_k)$.   The universal factors $W_{\alpha\beta}$ can be thought of as the effect of removing an M5-brane which fuses partitions $\alpha$ and $\beta$. It should be possible (see section \ref{wzw}) to obtain $W_{\alpha\beta}$ from the M2-brane perspective as some kind of partition function associated with the domain wall  represented by the M5-brane between the vacua labeled by $\alpha$ and $\beta$.  This should also be possible to do  using the ABJM theory.\\

The free energy $F=ln(Z)$ constructed from the partition function contains information about both the single particle and multi particle BPS states. The plethystic summation is used to project out the multiparticle states. Hence the function $F_{k_1k_2,...,k_n}$ counts single particle BPS states and can be expanded as
\bea
&F_{k_1,...,k_{N-1}}(\tau,m,\epsilon_1,\epsilon_2)=\nonumber\\&coefficient\hspace{0.1cm} of \hspace{0.1cm} Q^{k_1}_{f_1}...Q^{k_{N-1}}_{f_{N-1}}\hspace{0.1cm} in\hspace{0.1cm}\sum_{l\ge 1}\frac{\mu(l)}{l}log(Z_N(l\tau,lm,lt_{f_a},l\epsilon_1,l\epsilon_2))\nonumber\\
\eea

\subsection{Recursive structure  for the configuration of fully anti-symmetric Young diagrams}
Below we give  examples for a few configurations of the M5-M2 branes system . These examples show that there is no recursive structure for the full expressions of the free energies. Only a part of  the expression of the free energy shows the recursive structure. This part is what is alluded to before as 
\bea
Z_{\alpha_1\alpha_2...\alpha_k}+ all\hspace{0.1cm} possible\hspace{0.1cm} ways\hspace{0.1cm} of\hspace{0.1cm} factorizing
\eea

\subsection*{$\bullet$ $F_{2\emptyset\emptyset},F_{22\emptyset},F_{222}$}

\bea
&F_{2\emptyset\emptyset}(\tau,m,\epsilon_1,\epsilon_2)=-\frac{1}{2}Z_{1\emptyset\emptyset}(\tau,m,\epsilon_1,\epsilon_2)^2+\bigg[Z_{2\emptyset\emptyset}(\tau,m,\epsilon_1,\epsilon_2)\bigg]\nonumber\\&-\bigg(Z_{1\emptyset\emptyset}(2\tau,2m,2\epsilon_1,2\epsilon_2)\bigg)\nonumber\\
&F_{22\emptyset}(\tau,m,\epsilon_1,\epsilon_2)=
-\bigg[Z_{2\emptyset\emptyset}^2(\tau,m,\epsilon_1,\epsilon_2)-Z_{22\emptyset}(\tau,m,\epsilon_1,\epsilon_2)\bigg]-\nonumber\\&\bigg(F^{11\emptyset}(2\tau,2m,2\epsilon_1,2\epsilon_2)\bigg)+other\quad terms\nonumber\\
\eea
and  finally
\bea
&F_{222}(\tau,m,\epsilon_1,\epsilon_2)=\nonumber\\&
\bigg[Z_{2\emptyset\emptyset}^3(\tau,m,\epsilon_1,\epsilon_2) -2Z_{22\emptyset}(\tau,m,\epsilon_1,\epsilon_2) Z_{2\emptyset\emptyset}(\tau,m,\epsilon_1,\epsilon_2)\nonumber\\& +Z_{222}(\tau,m,\epsilon_1,\epsilon_2)\bigg] -\bigg(F_{111}(2\tau,2m,2\epsilon_1,2\epsilon_2) \bigg)+other\quad terms\nonumber\\
\eea
we now show that the terms in the square brackets  form a recursive structure. First we consider instanton number $k_i=2$
and the following Young diagrams
\bea
\nu_1=\{1,1\},\hspace{0.1cm} \nu_2=\{1,1\},...\hspace{0.1cm} and\hspace{0.1cm} so\hspace{0.1cm} on.
\eea
Using the notation $\theta_1(x\pm y):=\theta_1(x+ y)\theta_1(x- y)$ we find
\bea
&Z_{22\emptyset}(\tau,m,\epsilon_1,\epsilon_2)-Z_{2\emptyset\emptyset}^2(\tau,m,\epsilon_1,\epsilon_2)=Z_{2\emptyset\emptyset}(\tau,m,\epsilon_1,\epsilon_2)\mbox{\bf{W}}_2(\tau,m,\epsilon_1,\epsilon_2)\nonumber\\
\eea
where
\bea
&\mbox{\bf{W}}_2(\tau,m,\epsilon_1,\epsilon_2):=\frac{1}{{\theta_1(\epsilon_1-\epsilon_2)\theta_1(-2\epsilon_2)\theta_1(\epsilon_1)\theta_1(-\epsilon_2)}}\nonumber\\
&\times\big[\theta_1(-m\pm\frac{\epsilon_1}{2}\mp \frac{3\epsilon_2}{2})\theta_1(-m\pm \epsilon_-)-\theta_1(-m\pm \epsilon_+)\theta_1(-m\pm\frac{\epsilon_1}{2}\pm \frac{3\epsilon_2}{2})\big]\nonumber\\
\eea
To next order
\bea
&Z_{222}(\tau,m,\epsilon_1,\epsilon_2)+Z_{2\emptyset\emptyset}^3(\tau,m,\epsilon_1,\epsilon_2) -2Z_{22\emptyset}(\tau,m,\epsilon_1,\epsilon_2) Z_{2\emptyset\emptyset}(\tau,m,\epsilon_1,\epsilon_2)\nonumber\\&=
 Z_{2\emptyset\emptyset}(\tau,m,\epsilon_1,\epsilon_2)\mbox{\bf{W}}_2(\tau,m,\epsilon_1,\epsilon_2)^2\nonumber\\
\eea
\subsection*{$\bullet$ $F_{3\emptyset\emptyset},F_{33\emptyset},F_{333}$}
\bea
&F_{3\emptyset\emptyset}(\tau,m,\epsilon_1,\epsilon_2)=\bigg[Z_{3\emptyset\emptyset}(\tau,m,\epsilon_1,\epsilon_2)\bigg]-Z_{1\emptyset\emptyset}(\tau,m,\epsilon_1,\epsilon_2)Z_{2\emptyset\emptyset}(\tau,m,\epsilon_1,\epsilon_2)\nonumber\\&-\frac{1}{3}Z_{1\emptyset\emptyset}(\tau,m,\epsilon_1,\epsilon_2)^3\nonumber\\
\eea
\bea
&F_{33\emptyset}(\tau,m,\epsilon_1,\epsilon_2)=\bigg[Z_{33\emptyset}(\tau,m,\epsilon_1,\epsilon_2)-Z_{3\emptyset\emptyset}(\tau,m,\epsilon_1,\epsilon_2)^2\bigg]\nonumber\\&+\hspace{0.1cm} other\hspace{0.1cm} terms.
\eea
\bea
&F_{333}(\tau,m,\epsilon_1,\epsilon_2)=\bigg[Z_{333}(\tau,m,\epsilon_1,\epsilon_2)-2Z_{33\emptyset}(\tau,m,\epsilon_1,\epsilon_2)\nonumber\\ &Z_{3\emptyset\emptyset}(\tau,m,\epsilon_1,\epsilon_2)+
Z_{3\emptyset\emptyset}(\tau,m,\epsilon_1,\epsilon_2)^3\bigg]\nonumber\\&+\hspace{0.1cm} other\hspace{0.1cm} terms.
\eea
Now we consider the terms in square brackets for instanton number $k_i=3$ and the following Young diagrams
\bea
\nu_1=\{1,1,1\},\hspace{0.1cm} \nu_2=\{1,1,1\},...\hspace{0.1cm} and\hspace{0.1cm} so\hspace{0.1cm} on.
\eea
\bea
&Z_{33\emptyset}(\tau,m,\epsilon_1,\epsilon_2)-Z_{3\emptyset\emptyset}(\tau,m,\epsilon_1,\epsilon_2)^2\nonumber\\&=Z_{3\emptyset\emptyset}(\tau,m,\epsilon_1,\epsilon_2)\mbox{\bf{W}}_3(\tau,m,\epsilon_1,\epsilon_2)\nonumber\\
\eea
where
\bea
&\mbox{\bf{W}}_3(\tau,m,\epsilon_1,\epsilon_2)=\frac{1}{\theta_1(\epsilon_1-\epsilon_2)\theta_1(-2\epsilon_2)\theta_1(\epsilon_1)\theta_1(-\epsilon_2)\theta_1(-3\epsilon_2)\theta_1(\epsilon_1-2\epsilon_2)}\nonumber\\
&\times\big[\theta_1(-m\pm\frac{\epsilon_1}{2}\mp\frac{5\epsilon_2}{2})\theta_1(-m\pm\frac{\epsilon_1}{2}\mp\frac{3\epsilon_2}{2})\theta_1(-m\pm\epsilon_-)\nonumber\\&-\theta_1(-m\pm\frac{\epsilon_1}{2}\pm\frac{5\epsilon_2}{2})\theta_1(-m\pm\frac{\epsilon_1}{2}\pm\frac{3\epsilon_2}{2})\theta_1(-m\pm\epsilon_+) \big]\nonumber\\
\eea

To next order
\bea
&Z_{333}(\tau,m,\epsilon_1,\epsilon_2)-2Z_{33\emptyset}(\tau,m,\epsilon_1,\epsilon_2)Z_{3\emptyset\emptyset}(\tau,m,\epsilon_1,\epsilon_2)+\nonumber\\&
Z_{3\emptyset\emptyset}(\tau,m,\epsilon_1,\epsilon_2)^3=Z_{3\emptyset\emptyset}(\tau,m,\epsilon_1,\epsilon_2)\mbox{\bf{W}}_3(\tau,m,\epsilon_1,\epsilon_2)^2
\eea

\subsection{Observation}
We observe that the $W_i(\tau,m,\epsilon_1,\epsilon_2),i=1,...,N$ follow a pattern
\bea
&W_1(\tau,m,\epsilon_1,\epsilon_2)=\frac{\theta_1(-m\pm\epsilon_+)-\theta_1(-m\pm\epsilon_-)}{\theta_1(\epsilon_1)\theta_1(-\epsilon_2)}\nonumber\\
&W_2(\tau,m,\epsilon_1,\epsilon_2)
=\frac{\theta_1(-m\pm \epsilon_+)\theta_1(-m\pm\epsilon_+\pm\epsilon_2)-\theta_1(-m\pm \epsilon_-)\theta_1(-m\pm\epsilon_-\mp \epsilon_2)}{\theta_1(\epsilon_1)\theta_1(\epsilon_1-\epsilon_2)\theta_1(-\epsilon_2)\theta_1(-2\epsilon_2)}\nonumber\\
&W_3(\tau,m,\epsilon_1,\epsilon_2)
=\frac{\theta_1(-m\pm \epsilon_+)\theta_1(-m\pm\epsilon_+\pm\epsilon_2)\theta_1(-m\pm\epsilon_+\pm2\epsilon_2)}{\theta_1(\epsilon_1)\theta_1(\epsilon_1-\epsilon_2)\theta_1(\epsilon_1-2\epsilon_2)\theta_1(-\epsilon_2)\theta_1(-2\epsilon_2)\theta_1(-3\epsilon_2)}\nonumber\\ &-\frac{\theta_1(-m\pm \epsilon_-)\theta_1(-m\pm\epsilon_-\mp \epsilon_2)\theta_1(-m\pm\epsilon_-\mp 2\epsilon_2)}{\theta_1(\epsilon_1)\theta_1(\epsilon_1-\epsilon_2)\theta_1(\epsilon_1-2\epsilon_2)\theta_1(-\epsilon_2)\theta_1(-2\epsilon_2)\theta_1(-3\epsilon_2)}\nonumber\\
\eea
The above simple observation leads to the following generalisation 
\bea\label{eq:antisymmetricWN}
W_N(\tau,m,\epsilon_1,\epsilon_2)&=&\prod_{k=1}^N\bigg[\frac{\theta_1(-m\pm \epsilon_+\pm(k-1)\epsilon_2)}{\theta_1(-k\epsilon_2)\theta_1(\epsilon_1-(k-1)\epsilon_2)}\bigg]-\nonumber\\&&\prod_{k=1}^N\bigg[\frac{\theta_1(-m\pm \epsilon_-\mp(k-1)\epsilon_2)}{\theta_1(-k\epsilon_2)\theta_1(\epsilon_1-(k-1)\epsilon_2)}\bigg]\nonumber\\
\eea

It is curious to note that $W_N(\tau,m,\epsilon_1,\epsilon_2)$ can be written in terms of $W_1(\tau,m,\epsilon_1,\epsilon_2)$. We can rewrite $W_1(\tau,m,\epsilon_1,\epsilon_2)$ as
\bea
W_1(\tau,m,\epsilon_1,\epsilon_2)=\frac{\theta_1(-m\pm\epsilon_+)-\theta_1(-m\pm\epsilon_-)}{\theta_1(\epsilon_1)\theta_1(-\epsilon_2)}\nonumber\\:=W^{+}_1(\tau,-m\pm\epsilon_+,\epsilon_1,\epsilon_2)-W^{-}_1(\tau,-m\pm\epsilon_-,\epsilon_1,\epsilon_2)
\eea
Then $W_N(\tau,m,\epsilon_1,\epsilon_2)$ can be rewritten in terms of $W^{+}_1(\tau,-m\pm\epsilon_+,\epsilon_1,\epsilon_2)-$ and $W^{-}_1(\tau,-m\pm\epsilon_-,\epsilon_1,\epsilon_2)-$ as
\bea
&W_N(\tau,m,\epsilon_1,\epsilon_2)=\nonumber\\&\prod_{k=1}^N\bigg[W^{+}_1\bigg(\tau,-m\pm \epsilon_+\pm(k-1)\epsilon_2,-k\epsilon_2,\epsilon_1-(k-1)\epsilon_2\bigg)\bigg]-\nonumber\\&\prod_{k=1}^N\bigg[W_1^{-}\bigg(\tau,-m\pm \epsilon_-\mp(k-1)\epsilon_2,-k\epsilon_2,\epsilon_1-(k-1)\epsilon_2\bigg)\bigg]\nonumber\\
\eea
Note that under the modular transformation $(\tau,m,\epsilon_1,\epsilon_2)\to(-\frac{1}{\tau},\frac{m}{\tau},\frac{\epsilon_1}{\tau},\frac{\epsilon_2}{\tau})$, $W_N(\tau,m,\epsilon_1,\epsilon_2)$ transforms as
\bea
&W_N(-\frac{1}{\tau},\frac{m}{\tau},\frac{\epsilon_1}{\tau},\frac{\epsilon_2}{\tau})=\nonumber\\&\prod_{k=1}^Ne^{\frac{\pi i}{\tau} (2m^2+2(\epsilon_++(k-1)\epsilon_2)^2-k\epsilon_2^2)}\bigg[\frac{\theta_1(-m\pm \epsilon_+\pm(k-1)\epsilon_2)}{\theta_1(-k\epsilon_2)\theta_1(\epsilon_1-(k-1)\epsilon_2)}\bigg]\nonumber\\&-\prod_{k=1}^Ne^{\frac{\pi i}{\tau} (2m^2+2(\epsilon_--(k-1)\epsilon_2)^2-k\epsilon_2^2)}\bigg[\frac{\theta_1(-m\pm \epsilon_-\mp(k-1)\epsilon_2)}{\theta_1(-k\epsilon_2)\theta_1(\epsilon_1-(k-1)\epsilon_2)}\bigg]\nonumber\\
\eea
This shows that it is not modular covariant. However in the NS limit $\epsilon_2\to 0$ it becomes modular covariant.
For finite N, in the NS limit $\epsilon_1\to 0$, $W_N(\tau,m,\epsilon_2)$ reduces to the following expression
\bea
&W_N(\tau,m,\epsilon_2)^{NS}
= -\frac{\iota }{\eta(\tau)^3\theta_1(-\epsilon_2)\prod_{k=2}^N\theta_1(-k\epsilon_2)\theta_1(-(k-1)\epsilon_2)}\nonumber\\ &\times\bigg(\sum_{l=1}^N\theta_1^{\prime}(-m+\frac{2l-1}{2}\epsilon_2)\theta_1(-m-\frac{2l-1}{2}\epsilon_2)
\prod_{k=1,k\ne l}^N\theta_1(-m\pm\frac{2k-1}{2}\epsilon_2)\nonumber\\&-\sum_{l=1}^N\theta_1^{\prime}(-m-\frac{2l-1}{2}\epsilon_2)\theta_1(-m+\frac{2l-1}{2}\epsilon_2)
\prod_{k=1,k\ne l}^N\theta_1(-m\pm\frac{2k-1}{2}\epsilon_2)\bigg)\nonumber\\
\nonumber\\\nonumber\\
\eea
\subsection*{Multi-wrappings contribution}
if $n>1$ denotes the number of wrappings there are two choices\\
\subsection*{(a) $n$ contains repeated prime factors}
In this case there will be no contributions to the Free energy from multiwrappings.
\subsection*{(b) $n$ is equal to the product of $k$ \textit{distinct} prime factors}

For this case the generalised expression for  $W_N(\tau,m,\epsilon_1,\epsilon_2)$ is given as
\bea
&W_{n,N}(\tau,m,\epsilon_1,\epsilon_2)^{multi-wrappings}\equiv\prod_{k=1}^N\bigg[\frac{\theta_1(n\tau,-nm\pm n\epsilon_+\pm n(k-1)\epsilon_2)}{\theta_1(n\tau,-nk\epsilon_2)\theta_1(n\tau,n\epsilon_1-n(k-1)\epsilon_2)}\bigg]\nonumber\\&-\prod_{k=1}^N\bigg[\frac{\theta_1(n\tau,-nm\pm n\epsilon_-\mp n(k-1)\epsilon_2)}{\theta_1(n\tau,-nk\epsilon_2)\theta_1(n\tau,n\epsilon_1-n(k-1)\epsilon_2)}\bigg]\nonumber\\&=W_N(n\tau,nm,n\epsilon_1,n\epsilon_2)\nonumber
\eea
This result confirms that fact  that the correct objects to decompose for multiple M2-branes configurations are the components  $Z_{\nu_1\nu_2...\nu_n}$ and not the free energies or BPS degeneracies $F$'s.
\subsection{Mixed Partitions}

\subsection*{$\bullet$ $F_{12\emptyset\emptyset\emptyset\emptyset},F_{1212\emptyset\emptyset},F_{121212}$}

\bea
&Z_{1212\emptyset\emptyset}(\tau,m,\epsilon_1,\epsilon_2)-Z_{12\emptyset\emptyset\emptyset\emptyset}^2(\tau,m,\epsilon_1,\epsilon_2)=\nonumber\\&Z_{12\emptyset\emptyset\emptyset\emptyset}(\tau,m,\epsilon_1,\epsilon_2)\mbox{\bf{W}}_{12}(\tau,m,\epsilon_1,\epsilon_2)
\eea
where
\bea
&\mbox{\bf{W}}_{12}(\tau,m,\epsilon_1,\epsilon_2)=\frac{1}{D_{12\emptyset\emptyset\emptyset\emptyset}(\tau,m,\epsilon_1,\epsilon_2)}\times\theta _1\left(\frac{1}{2} \left(-2 m+\epsilon _1-3 \epsilon _2\right)\right) \theta
   _1\left(-m-\frac{\epsilon _1}{2}-\frac{\epsilon _2}{2}\right)
\nonumber\\&   \theta
   _1\left(\frac{1}{2} \left(-2 m-\epsilon _1+\epsilon _2\right)\right) \theta
   _1\left(\frac{1}{2} \left(-2 m+\epsilon _1+\epsilon _2\right)\right)\times \bigg[\theta
   _1\left(\frac{1}{2} \left(-2 m+\epsilon _1-\epsilon _2\right)\right)\nonumber\\& \theta
   _1\left(-m-\frac{\epsilon _1}{2}+\frac{3 \epsilon _2}{2}\right)
   -\theta
   _1\left(-m-\frac{\epsilon _1}{2}-\frac{\epsilon _2}{2}\right) \theta
   _1\left(\frac{1}{2} \left(-2 m+\epsilon _1+3 \epsilon _2\right)\right)\bigg]\nonumber\\
 \eea
 \subsection*{$\bullet$ $Z_{\nu_1\nu_2\nu_3}$\quad for\quad $\nu_1=\{1,1,1\},\nu_2=\{3\},\nu_3=\{2,1\}$}
 A non-trivial example of  mixed Young diagrams case is 
 $Z_{\nu_1\nu_2\nu_3}$ where $\nu_1$ is the fully symmetric Young diagram, $\nu_2$ is the fully antisymmetric Young diagram, and $\nu_3=\{2,1\}$. Note that $|\nu_1|=|\nu_2|=|\nu_3|=3$ In this case we get
 \bea
 &Z_{\nu_1\nu_2\nu_3}(\tau,m,\epsilon_1,\epsilon_2)-Z_{\nu_1}(\tau,m,\epsilon_1,\epsilon_2)Z_{\nu_2\nu_3}(\tau,m,\epsilon_1,\epsilon_2)\nonumber\\&-Z_{\nu_1\nu_2}(\tau,m,\epsilon_1,\epsilon_2)Z_{\nu_3}(\tau,m,\epsilon_1,\epsilon_2)+\nonumber\\&Z_{\nu_1}(\tau,m,\epsilon_1,\epsilon_2)Z_{\nu_2}(\tau,m,\epsilon_1,\epsilon_2)Z_{\nu_2}(\tau,m,\epsilon_1,\epsilon_2)=\nonumber\\&\frac{1}{\theta_1(-2\epsilon_1)\theta_1(-\epsilon_1)^2\theta_1(\epsilon_1)^2\theta_1(3\epsilon_1)\theta_1(\epsilon_1-2\epsilon_2)^2\theta_1(-\epsilon_2+2\epsilon_1)^2\theta_1(-3\epsilon_2)}\nonumber\\&\times\frac{1}{\theta_1(-\epsilon_2)^2\theta_1(\epsilon_2)^2\theta_1(2\epsilon_2)\theta_1(\epsilon_2-\epsilon_1)^2}\nonumber\\&
 \theta_1(\frac{1}{2}(\epsilon_1-3\epsilon_2-2m))\theta_1(\frac{1}{2}(\epsilon_1-\epsilon_2-2m))\theta_1(\frac{1}{2}(\epsilon_1+\epsilon_2-2m))^3\nonumber\\&\theta_1(\frac{1}{2}(3\epsilon_1+\epsilon_2-2m))\theta_1(\frac{1}{2}(-\epsilon_1+3\epsilon_2-2m))\theta_1(\frac{1}{2}(\epsilon_1+3\epsilon_2-2m))\nonumber\\&\bigg[\theta_1(\frac{1}{2}(5\epsilon_1-3\epsilon_2-2m))\theta_1(\frac{1}{2}(-\epsilon_1-\epsilon_2-2m))\theta_1(\frac{1}{2}(-\epsilon_1+\epsilon_2-2m))^2-\nonumber\\&\theta_1(\frac{1}{2}(3\epsilon_1+\epsilon_2-2m))\theta_1(\frac{1}{2}(5\epsilon_1+\epsilon_2-2m))\theta_1(\frac{1}{2}(-\epsilon_1+5\epsilon_2-2m))\nonumber\\&\theta_1(\frac{1}{2}(\epsilon_1+3\epsilon_2-2m)) \bigg]\nonumber\\&
 \bigg[\theta_1(\frac{1}{2}(\epsilon_1-\epsilon_2-2m))\theta_1(\frac{1}{2}(-3\epsilon_1-\epsilon_2-2m))^2\theta_1(\frac{1}{2}(-\epsilon_1-\epsilon_2-2m))^2\nonumber\\&\theta_1(\frac{1}{2}(-5\epsilon_1+5\epsilon_2-2m))-\nonumber\\&\theta_1(\frac{1}{2}(3\epsilon_1-\epsilon_2-2m))\theta_1(\frac{1}{2}(5\epsilon_1-\epsilon_2-2m))\theta_1(\frac{1}{2}(-3\epsilon_1+\epsilon_2-2m))\nonumber\\&\theta_1(\frac{1}{2}(5\epsilon_1-\epsilon_2-2m))\theta_1(\frac{1}{2}(-3\epsilon_1+\epsilon_2-2m))\theta_1(\frac{1}{2}(-\epsilon_1+\epsilon_2-2m))\nonumber\\&\theta_1(\frac{1}{2}(\epsilon_1+\epsilon_2-2m))\theta_1(\frac{1}{2}(\epsilon_1+5\epsilon_2-2m)) \bigg]
 \eea
 
  \subsection*{$\bullet$ $Z_{\nu_1\nu_2\nu_3}$\quad for\quad $\nu_1=\{2,1,1\},\nu_2=\{3,1\},\nu_3=\{2,2\}$}
  A second non-trivial example of  mixed Young diagrams corresponds to $\nu_1=\{2,1,1\},\nu_2=\{3,1\},\nu_3=\{2,2\}$. Note that $|\nu_1|=|\nu_2|=|\nu_3|=4$ in this case. We get
   \bea
 &Z_{\nu_1\nu_2\nu_3}(\tau,m,\epsilon_1,\epsilon_2)-Z_{\nu_1}(\tau,m,\epsilon_1,\epsilon_2)Z_{\nu_2\nu_3}(\tau,m,\epsilon_1,\epsilon_2)\nonumber\\&-Z_{\nu_1\nu_2}(\tau,m,\epsilon_1,\epsilon_2)Z_{\nu_3}(\tau,m,\epsilon_1,\epsilon_2)+\nonumber\\&Z_{\nu_1}(\tau,m,\epsilon_1,\epsilon_2)Z_{\nu_2}(\tau,m,\epsilon_1,\epsilon_2)Z_{\nu_2}(\tau,m,\epsilon_1,\epsilon_2)=\nonumber\\&\frac{1}{\theta_1(-\epsilon_1)^2\theta_1(\epsilon_1)^4\theta_1(2\epsilon_1)\theta_1(\epsilon_1-3\epsilon_2)\theta_1(\epsilon_1-2\epsilon_2)\theta_1(2\epsilon_1-2\epsilon_2)^2\theta_1(\epsilon_1-\epsilon_2)^3}\nonumber\\&\times\frac{1}{\theta_1(2\epsilon_1-\epsilon_2)\theta_1(3\epsilon_1-\epsilon_2)\theta_1(-2\epsilon_2)^2\theta_1(-\epsilon_2)^4)\theta_1(\epsilon_2)^2}\nonumber\\&
 \theta_1(\frac{1}{2}(\epsilon_1-\epsilon_2-2m))\theta_1(\frac{1}{2}(\epsilon_1+\epsilon_2-2m))^3\theta_1(\frac{1}{2}(3\epsilon_1+\epsilon_2-2m))^2\nonumber\\&\theta_1(\frac{1}{2}(\epsilon_1+3\epsilon_2-2m))^2\theta_1(\frac{1}{2}(3\epsilon_1+3\epsilon_2-2m))\nonumber\\&\bigg[\theta_1(\frac{1}{2}(3\epsilon_1-3\epsilon_2-2m))\theta_1(\frac{1}{2}(\epsilon_1-\epsilon_2-2m))^2\theta_1(\frac{1}{2}(-\epsilon_1-\epsilon_2-2m))^2\nonumber\\&\theta_1(\frac{1}{2}(-3\epsilon_1+\epsilon_2-2m))^2\theta_1(\frac{1}{2}(-5\epsilon_1+5\epsilon_2-2m))-\nonumber\\&\theta_1(\frac{1}{2}(5\epsilon_1-\epsilon_2-2m))\theta_1(\frac{1}{2}(-\epsilon_1+\epsilon_2-2m))\theta_1(\frac{1}{2}(\epsilon_1+\epsilon_2-2m))\nonumber\\&\theta_1(\frac{1}{2}(3\epsilon_1+\epsilon_2-2m))^3\theta_1(\frac{1}{2}(\epsilon_1+3\epsilon_2-2m))\theta_1(\frac{1}{2}(\epsilon_1+5\epsilon_2-2m)) \bigg]\nonumber\\&
 \bigg[\theta_1(\frac{1}{2}(5\epsilon_1-3\epsilon_2-2m))\theta_1(\frac{1}{2}(\epsilon_1-\epsilon_2-2m))^2\theta_1(\frac{1}{2}(-\epsilon_1+\epsilon_2-2m))^2\nonumber\\&\theta_1(\frac{1}{2}(-3\epsilon_1+3\epsilon_2-2m))\theta_1(\frac{1}{2}(-\epsilon_1+3\epsilon_2-2m))-\nonumber\\&\theta_1(\frac{1}{2}(3\epsilon_1+\epsilon_2-2m))\theta_1(\frac{1}{2}(5\epsilon_1+\epsilon_2-2m))\theta_1(\frac{1}{2}(-\epsilon_1+5\epsilon_2-2m))\nonumber\\&\theta_1(\frac{1}{2}(\epsilon_1+3\epsilon_2-2m))^3\theta_1(\frac{1}{2}(3\epsilon_1+3\epsilon_2-2m)) \bigg]
 \eea
 
 This is in line with the statement given in (\ref{eq:fusion}) that  if there are $k$ partitions $\alpha_1,...,\alpha_k$ of the same size .i.e. $|\alpha_1|=...=|\nu_k|=n$ then it is the case that:
\bea
Z_{\alpha_1\alpha_2...\alpha_k}&=& all\hspace{0.1cm} possible\hspace{0.1cm} ways\hspace{0.1cm} of\hspace{0.1cm} factorizing\nonumber\\&+& Z_{\beta} W_{\alpha_1\alpha_2} W_{\alpha_2\alpha_3}....W_{\alpha_{k-2}\alpha_{k-1}}W_{\alpha_{k-1}\alpha_{k}}
\eea
where $\beta$ is the result of  fusing \cite{Gaiotto:2008ak}  the partitions $(\alpha_1,\alpha_2,...,\alpha_k)$.

  \subsection*{General configuration}
In the general configuration we consider $r+1$ M5-branes with $k_i,i=1,...,r$ M2-branes between $i$-th and $(i+1)$-th M5-branes. The Young diagrams labelling the M2-branes are in antisymmetric representations. The corresponding components of the elliptic genus $Z_{k_1k_2...k_r}(\tau,m,\epsilon_1,\epsilon_2)$ have following recursive pattern

\bea\label{eq:genconf}
&Z_{k_1k_2...k_rk_1k_2...k_rk_1k_2...k_r}(\tau,m,\epsilon_1,\epsilon_2)-2Z_{k_1k_2...k_rk_1k_2...k_r\emptyset\emptyset...\emptyset}(\tau,m,\epsilon_1,\epsilon_2)\nonumber\\&Z_{k_1k_2...k_r\emptyset\emptyset...\emptyset\emptyset\emptyset...\emptyset}(\tau,m,\epsilon_1,\epsilon_2)+Z_{k_1k_2...k_r\emptyset\emptyset...\emptyset\emptyset\emptyset...\emptyset}^3(\tau,m,\epsilon_1,\epsilon_2) =\nonumber\\&
\hspace{0.5cm}
\Bigg\{\Big[\prod_{l=1}^{k_1}\theta_1(-m+\frac{\epsilon_1}{2}-\epsilon_2(k_2-l+\frac{1}{2}))\theta_1(-m-\frac{\epsilon_1}{2}-\epsilon_2(l-\frac{1}{2}))\nonumber\\&
\prod_{l=1}^{k_2}\theta_1(-m+\frac{\epsilon_1}{2}-\epsilon_2(k_3-l+\frac{1}{2}))\theta_1(-m-\frac{\epsilon_1}{2}-\epsilon_2(-k_1+l-\frac{1}{2}))\nonumber\\&
\hspace{0.5cm}.............\nonumber\\&
\prod_{l=1}^{k_r}\theta_1(-m+\frac{\epsilon_1}{2}-\epsilon_2(k_1-l+\frac{1}{2}))\theta_1(-m-\frac{\epsilon_1}{2}-\epsilon_2(-k_{r-1}+l-\frac{1}{2}))\Big]\nonumber\\&\times
\Big[\prod_{l=1}^{k_1}\theta_1(-m+\frac{\epsilon_1}{2}-\epsilon_2(k_2-l+\frac{1}{2}))\theta_1(-m-\frac{\epsilon_1}{2}-\epsilon_2(-k_r+l-\frac{1}{2}))\nonumber\\&
\prod_{l=1}^{k_2}\theta_1(-m+\frac{\epsilon_1}{2}-\epsilon_2(k_3-l+\frac{1}{2}))\theta_1(-m-\frac{\epsilon_1}{2}-\epsilon_2(-k_1+l-\frac{1}{2}))\nonumber\\&
\hspace{0.5cm}.............\nonumber\\&
\prod_{l=1}^{k_r}\theta_1(-m+\frac{\epsilon_1}{2}-\epsilon_2(k_1-l+\frac{1}{2}))\theta_1(-m-\frac{\epsilon_1}{2}-\epsilon_2(-k_{r-1}+l-\frac{1}{2}))\Big]\nonumber\\&\times
\hspace{0.5cm}\bigg[\prod_{l=1}^{k_1}\theta_1(-m+\frac{\epsilon_1}{2}-\epsilon_2(k_2-l+\frac{1}{2}))\theta_1(-m-\frac{\epsilon_1}{2}-\epsilon_2(-k_r+l-\frac{1}{2}))\nonumber\\&
\prod_{l=1}^{k_2}\theta_1(-m+\frac{\epsilon_1}{2}-\epsilon_2(k_3-l+\frac{1}{2}))\theta_1(-m-\frac{\epsilon_1}{2}-\epsilon_2(-k_1+l-\frac{1}{2}))\nonumber\\&
\hspace{0.5cm}.............\nonumber\\&
\prod_{l=1}^{k_r}\theta_1(-m+\frac{\epsilon_1}{2}-\epsilon_2(-l+\frac{1}{2}))\theta_1(-m-\frac{\epsilon_1}{2}-\epsilon_2(-k_{r-1}+l-\frac{1}{2}))\bigg]\Bigg\}\nonumber\\&
-2\Bigg\{\bigg[\prod_{l=1}^{k_1}\theta_1(-m+\frac{\epsilon_1}{2}-\epsilon_2(k_2-l+\frac{1}{2}))\theta_1(-m-\frac{\epsilon_1}{2}-\epsilon_2(l-\frac{1}{2}))\nonumber\\&
\prod_{l=1}^{k_2}\theta_1(-m+\frac{\epsilon_1}{2}-\epsilon_2(k_3-l+\frac{1}{2}))\theta_1(-m-\frac{\epsilon_1}{2}-\epsilon_2(-k_1+l-\frac{1}{2}))\nonumber\\&
\hspace{0.5cm}.............\nonumber\\&
\prod_{l=1}^{k_r}\theta_1(-m+\frac{\epsilon_1}{2}-\epsilon_2(k_1-l+\frac{1}{2}))\theta_1(-m-\frac{\epsilon_1}{2}-\epsilon_2(-k_{r-1}+l-\frac{1}{2}))\nonumber\\&
\prod_{l=1}^{k_1}\theta_1(-m+\frac{\epsilon_1}{2}-\epsilon_2(k_2-l+\frac{1}{2}))\theta_1(-m-\frac{\epsilon_1}{2}-\epsilon_2(-k_r+l-\frac{1}{2}))\nonumber\\&
\prod_{l=1}^{k_2}\theta_1(-m+\frac{\epsilon_1}{2}-\epsilon_2(k_3-l+\frac{1}{2}))\theta_1(-m-\frac{\epsilon_1}{2}-\epsilon_2(-k_1+l-\frac{1}{2}))\nonumber\\&
\hspace{0.5cm}.............\nonumber\\&
\prod_{l=1}^{k_r}\theta_1(-m+\frac{\epsilon_1}{2}-\epsilon_2(-l+\frac{1}{2}))\theta_1(-m-\frac{\epsilon_1}{2}-\epsilon_2(-k_{r-1}+l-\frac{1}{2}))\bigg]\nonumber\\&
\bigg[\prod_{l=1}^{k_1}\theta_1(-m+\frac{\epsilon_1}{2}-\epsilon_2(k_2-l+\frac{1}{2}))\theta_1(-m-\frac{\epsilon_1}{2}-\epsilon_2(l-\frac{1}{2}))\nonumber\\&
\prod_{l=1}^{k_2}\theta_1(-m+\frac{\epsilon_1}{2}-\epsilon_2(k_3-l+\frac{1}{2}))\theta_1(-m-\frac{\epsilon_1}{2}-\epsilon_2(-k_1+l-\frac{1}{2}))\nonumber\\&
\hspace{0.5cm}.............\nonumber\\&
\prod_{l=1}^{k_r}\theta_1(-m+\frac{\epsilon_1}{2}-\epsilon_2(-l+\frac{1}{2}))\theta_1(-m-\frac{\epsilon_1}{2}-\epsilon_2(-k_{r-1}+l-\frac{1}{2}))\bigg]\Bigg\}\nonumber
\eea
\bea\label{eq:generalrecursive}
&+\Bigg\{\prod_{l=1}^{k_1}\theta_1(-m+\frac{\epsilon_1}{2}-\epsilon_2(k_2-l+\frac{1}{2}))\theta_1(-m-\frac{\epsilon_1}{2}-\epsilon_2(l-\frac{1}{2}))\nonumber\\&
\prod_{l=1}^{k_2}\theta_1(-m+\frac{\epsilon_1}{2}-\epsilon_2(k_3-l+\frac{1}{2}))\theta_1(-m-\frac{\epsilon_1}{2}-\epsilon_2(-k_1+l-\frac{1}{2}))\nonumber\\&
\hspace{0.5cm}.............\nonumber\\&
\prod_{l=1}^{k_r}\theta_1(-m+\frac{\epsilon_1}{2}-\epsilon_2(-l+\frac{1}{2}))\theta_1(-m-\frac{\epsilon_1}{2}-\epsilon_2(-k_{r-1}+l-\frac{1}{2}))\nonumber\\&\Bigg\}^3\nonumber\\&
=\Bigg\{\prod_{l=1}^{k_1}\theta_1(-m+\frac{\epsilon_1}{2}-\epsilon_2(k_2-l+\frac{1}{2}))\theta_1(-m-\frac{\epsilon_1}{2}-\epsilon_2(l-\frac{1}{2}))\nonumber\\&
\prod_{l=1}^{k_2}\theta_1(-m+\frac{\epsilon_1}{2}-\epsilon_2(k_3-l+\frac{1}{2}))\theta_1(-m-\frac{\epsilon_1}{2}-\epsilon_2(-k_1+l-\frac{1}{2}))\nonumber\\&
\hspace{0.5cm}.............\nonumber\\&
\prod_{l=1}^{k_r}\theta_1(-m+\frac{\epsilon_1}{2}-\epsilon_2(-l+\frac{1}{2}))\theta_1(-m-\frac{\epsilon_1}{2}-\epsilon_2(-k_{r-1}+l-\frac{1}{2}))\Bigg\}\nonumber\\&
\Bigg\{\Bigg(\nonumber\\&\prod_{l=1}^{k_r}\theta_1(-m+\frac{\epsilon_1}{2}-\epsilon_2(k_1-l+\frac{1}{2}))\theta_1(-m-\frac{\epsilon_1}{2}-\epsilon_2(-k_{r-1}+l-\frac{1}{2}))\nonumber\\&\times
\prod_{l=1}^{k_1}\theta_1(-m+\frac{\epsilon_1}{2}-\epsilon_2(k_2-l+\frac{1}{2}))\theta_1(-m-\frac{\epsilon_1}{2}-\epsilon_2(-k_r+l-\frac{1}{2}))\nonumber\\&
\prod_{l=1}^{k_2}\theta_1(-m+\frac{\epsilon_1}{2}-\epsilon_2(k_3-l+\frac{1}{2}))\theta_1(-m-\frac{\epsilon_1}{2}-\epsilon_2(-k_1+l-\frac{1}{2}))\nonumber\\&
\hspace{0.5cm}.............\nonumber\\&
\prod_{l=1}^{k_{r-1}}\theta_1(-m+\frac{\epsilon_1}{2}-\epsilon_2(k_r-l+\frac{1}{2}))\nonumber\\&\theta_1(-m-\frac{\epsilon_1}{2}-\epsilon_2(-k_{r-2}+l-\frac{1}{2}))\Bigg)^2\nonumber\\&
-2\Bigg[\prod_{l=1}^{k_1}\theta_1(-m+\frac{\epsilon_1}{2}-\epsilon_2(k_2-l+\frac{1}{2}))\theta_1(-m-\frac{\epsilon_1}{2}-\epsilon_2(l-\frac{1}{2}))\nonumber\\&
\prod_{l=1}^{k_2}\theta_1(-m+\frac{\epsilon_1}{2}-\epsilon_2(k_3-l+\frac{1}{2}))\theta_1(-m-\frac{\epsilon_1}{2}-\epsilon_2(-k_1+l-\frac{1}{2}))\nonumber\\&
\hspace{0.5cm}.............\nonumber\\&
\prod_{l=1}^{k_r}\theta_1(-m+\frac{\epsilon_1}{2}-\epsilon_2(k_1-l+\frac{1}{2}))\theta_1(-m-\frac{\epsilon_1}{2}-\epsilon_2(-k_{r-1}+l-\frac{1}{2}))\nonumber\\&
\prod_{l=1}^{k_1}\theta_1(-m+\frac{\epsilon_1}{2}-\epsilon_2(k_2-l+\frac{1}{2}))\theta_1(-m-\frac{\epsilon_1}{2}-\epsilon_2(-k_r+l-\frac{1}{2}))\nonumber\\&
\prod_{l=1}^{k_2}\theta_1(-m+\frac{\epsilon_1}{2}-\epsilon_2(k_3-l+\frac{1}{2}))\theta_1(-m-\frac{\epsilon_1}{2}-\epsilon_2(-k_1+l-\frac{1}{2}))\nonumber\\&
\hspace{0.5cm}.............\nonumber\\&
\prod_{l=1}^{k_r}\theta_1(-m+\frac{\epsilon_1}{2}-\epsilon_2(-l+\frac{1}{2}))\theta_1(-m-\frac{\epsilon_1}{2}-\epsilon_2(-k_{r-1}+l-\frac{1}{2}))\Bigg]\nonumber
\eea
\bea
&+\Bigg(\prod_{l=1}^{k_1}\theta_1(-m+\frac{\epsilon_1}{2}-\epsilon_2(k_2-l+\frac{1}{2}))\theta_1(-m-\frac{\epsilon_1}{2}-\epsilon_2(l-\frac{1}{2}))\nonumber\\&
\prod_{l=1}^{k_2}\theta_1(-m+\frac{\epsilon_1}{2}-\epsilon_2(k_3-l+\frac{1}{2}))\theta_1(-m-\frac{\epsilon_1}{2}-\epsilon_2(-k_1+l-\frac{1}{2}))\nonumber\\&
\hspace{0.5cm}.............\nonumber\\&
\prod_{l=1}^{k_r}\theta_1(-m+\frac{\epsilon_1}{2}-\epsilon_2(-l+\frac{1}{2}))\theta_1(-m-\frac{\epsilon_1}{2}-\epsilon_2(-k_{r-1}+l-\frac{1}{2}))\nonumber\\&\Bigg)^2\Bigg\}\nonumber
\eea
\bea
&=Z_{k_1k_2...k_r\emptyset\emptyset...\emptyset\emptyset...\emptyset}\Bigg\{\prod_{l=1}^{k_1}\theta_1(-m+\frac{\epsilon_1}{2}-\epsilon_2(k_2-l+\frac{1}{2}))\nonumber\\&\theta_1(-m-\frac{\epsilon_1}{2}-\epsilon_2(l-\frac{1}{2}))\nonumber\\&
\prod_{l=1}^{k_2}\theta_1(-m+\frac{\epsilon_1}{2}-\epsilon_2(k_3-l+\frac{1}{2}))\theta_1(-m-\frac{\epsilon_1}{2}-\epsilon_2(-k_1+l-\frac{1}{2}))\nonumber\\&
\hspace{0.5cm}.............\nonumber\\&
\prod_{l=1}^{k_r}\theta_1(-m+\frac{\epsilon_1}{2}-\epsilon_2(-l+\frac{1}{2}))\theta_1(-m-\frac{\epsilon_1}{2}-\epsilon_2(-k_{r-1}+l-\frac{1}{2}))\nonumber\\&-\vspace{0.5cm}\nonumber\\&\prod_{l=1}^{k_1}\theta_1(-m+\frac{\epsilon_1}{2}-\epsilon_2(k_2-l+\frac{1}{2}))\theta_1(-m-\frac{\epsilon_1}{2}-\epsilon_2(l-\frac{1}{2}))\nonumber\\&
\prod_{l=1}^{k_2}\theta_1(-m+\frac{\epsilon_1}{2}-\epsilon_2(k_3-l+\frac{1}{2}))\theta_1(-m-\frac{\epsilon_1}{2}-\epsilon_2(-k_1+l-\frac{1}{2}))\nonumber\\&
\hspace{0.5cm}.............\nonumber\\&
\prod_{l=1}^{k_r}\theta_1(-m+\frac{\epsilon_1}{2}-\epsilon_2(-l+\frac{1}{2}))\theta_1(-m-\frac{\epsilon_1}{2}-\epsilon_2(-k_{r-1}+l-\frac{1}{2}))\nonumber\\&
\Bigg\}^2\nonumber\\&\vspace{0.5cm}\nonumber\\&
=Z_{k_1k_2...k_r\emptyset \emptyset...\emptyset\emptyset...\emptyset}(\tau,m,\epsilon_1,\epsilon_2)\mbox{\bf{W}}_{k_1k_2...k_r}^2(\tau,m,\epsilon_1,\epsilon_2)
\eea

\subsection{Fully symmetric configuration of Young diagrams}
The symmetric configuration of Young diagram is the complex conjugate of the antisymmetric configuration and this amounts to replacing $\epsilon_1$ with $\epsilon_2$ and vice versa. 
The results obtained in the previous section remains valid in this situation if we make the replacement
\bea
\epsilon_2\leftrightarrow \epsilon_1
\eea
\bea\label{eq:symmetricWN}
&W_N(\tau,m,\epsilon_1,\epsilon_2)=\prod_{k=1}^N\bigg[\frac{\theta_1(-m\pm \epsilon_+\pm(k-1)\epsilon_1)}{\theta_1(-k\epsilon_1)\theta_1(\epsilon_2-(k-1)\epsilon_1)}\bigg]\nonumber\\&-\prod_{k=1}^N\bigg[\frac{\theta_1(-m\pm \epsilon_-\mp(k-1)\epsilon_1)}{\theta_1(-k\epsilon_1)\theta_1(\epsilon_2-(k-1)\epsilon_1)}\bigg]\nonumber\\
\eea
For finite N, in the NS limit $\epsilon_2\to 0$ we get
\bea
&W_N(\tau,m,\epsilon_2)^{NS}
= -\frac{\iota }{\eta(\tau)^3\theta_1(-\epsilon_1)\prod_{k=2}^N\theta_1(-k\epsilon_1)\theta_1(-(k-1)\epsilon_1)}\nonumber\\ &\times\bigg(\sum_{l=1}^N\theta_1^{\prime}(-m+\frac{2l-1}{2}\epsilon_1)\theta_1(-m-\frac{2l-1}{2}\epsilon_1)\nonumber\\&
\prod_{k=1,k\ne l}^N\theta_1(-m\pm\frac{2k-1}{2}\epsilon_1)\nonumber\\&-\sum_{l=1}^N\theta_1^{\prime}(-m-\frac{2l-1}{2}\epsilon_1)\theta_1(-m+\frac{2l-1}{2}\epsilon_1)\nonumber\\&
\prod_{k=1,k\ne l}^N\theta_1(-m\pm\frac{2k-1}{2}\epsilon_1)\bigg)\nonumber
\eea

\subsection{Remark : Some comments on the algebra of holomorphic curves and the recursive structure }
Gopakumar and Vafa  reformulated \cite{Gopakumar:1998ii,Gopakumar:1998jq} the topological string  amplitudes focussing on the target space perspective. The 5d $N=1^*$ supersymmetric gauge theory, for a given M-theory CY3-fold compactification, has
BPS particles. These BPS particles correspond to the M2-branes wrapped on holomorphic curves in the CY3-fold. The quantum numbers of the BPS particles are given by the curve class $\Sigma\in H_2(CY3,\mathbb{Z})$
 and the spin content of the 5d little group of the massive particles $(j_R,j_L)=SU(2)_R\times SU(2)_L$. The particle content with charge $\Sigma$ and spins $(j_R,j_L)$ is an invariant for a non-compact CY3-fold and is denoted by $N_{\Sigma}^{(j_R,j_L)}$. The moduli space furnished by the D2-branes wrapped on $\Sigma$ is topologically non-trivial and  the number of its cohomology classes is equal to $N_{\Sigma}^{(j_R,j_L)}$.  
The explicit form of the topological string partition function in given by
\bea
Z(\omega,\epsilon)&=&\prod_{\Sigma\in H_2(CY3)}\prod_{j_L}\nonumber\\&\times&\prod_{k=-j_L}^{j_L}\prod_{m=0}^{\infty}(1-q^{2k+m+1}Q^{\Sigma})^{(-1)^{2j_L+1}(m+1)N_{\Sigma}^{(j_R,j_L)}}\nonumber\\
\eea
where $Q^{\Sigma}=e^{-T_{\Sigma}}$, $q=e^{-i\epsilon}$ and $\omega$ is the K\"ahler form on the CY3-fold. It was shown in \cite{Gopakumar:1998ii} that for $j_L=0$ the partition function $Z(\omega,\epsilon)$ counts the states in a Hilbert space. It is interesting to note that for a given particle content $N_{\Sigma}^{(j_R,j_L)}$, the above partition function $Z(\omega,\epsilon)$  can be written as an index as \cite{Hollowood:2003cv}
\bea
Z=Tr_{\mathcal{H}}(-1)^{2(j_L+j_R)}q^{2j_L^3}e^{-T_{\Sigma}}
\eea
where $\mathcal{H}$ is the quantized Hilbert subspace containing holomorphic modes of the BPS fields and $T_{\Sigma}$ is the Hamiltonian of the theory. In other words the topological string partition function can be interpreted  as the counting  the  holomorphic (components of the) BPS states in the quantized Hilbert space.
\\
Moreover as shown in section (\ref{recursivestructure}) in the computation of the M-strings  partition function 
one has to use fixed point theorems. For that purpose it is necessary to determine the equivariant weights of certain vector bundles on the M-strings moduli space. In the description of the vector bundles the $\mbox{Ext}$-groups  make an appearance. These groups appear  \cite{Sharpe:2003dr} in the counting problem of open string states between the D-branes wrapped on the holomorphic submanifolds. \\
 It was shown in \cite{Hohenegger:2015btj} that the free energies $F^{1_1,1_2,...,1_N}(\tau,m,\epsilon_1,\epsilon_2)$ for a configuration of  finitely separated $N+1$ M5-branes with a single M2-brane  stretched  between consecutive M5-branes, is reducible and recursive such that 
\bea\label{selfsim}
F^{1_1,1_2,...,1_N}(\tau,m,\epsilon_1,\epsilon_2)=W(\tau,m,\epsilon_1,\epsilon_2)^{N-1}F^{1_1,0,...,0}(\tau,m,\epsilon_1,\epsilon_2)
\eea
This shows that  the factor of  $W(\tau,m,\epsilon_1,\epsilon_2)$ appears  everytime a single M5-brane is removed from the configuration $--M2-M5-M2---$. This recursive structure, as shown in \cite{Hohenegger:2016eqy}, indicates that the degrees of freedom corresponding to the  $F^{1_1,1_2,...,1_N}(\tau,m,\epsilon_1,\epsilon_2)$ can be obtained from $F^{1_1,0,...,0}(\tau,m,\epsilon_1,\epsilon_2)$ upto the $univeral$ factor  $W(\tau,m,\epsilon_1,\epsilon_2)$. A similar interpretation is expected from the generalised $W_N(\tau,m,\epsilon_1,\epsilon_2)$ computed in section (\ref{recursivestructure}).\\
The M-string configuration when lifted to the higher dimensional F-theory corresponds \cite{DelZotto:2014hpa} to an elliptic CY3-fold in which a $D3$-brane wraps a $\mathbb{P}^1$ whose normal bundle is $\mathcal{O}(-2)$. The configuration of multiple parallel M5-branes with M2-branes stretching between them corresponds to D3-branes wrapping a chain of $\mathbb{P}^1$s with $\mathcal{O}(-2)$ normal bundles. The elliptic CY3-fold is a resolved $A_{N-1}$ fibration over the $\mbox{T}^2$.  In this set up the M5-branes correspond to a holomorphic $(-2)$ curve. In the case under consideration the holomorphic cycles correspond to the positive roots of  the gauge group $SU(N)$ of the $5d,\mathcal{N}=1^*$ gauge theory. Recalling that  \cite{Morrison:2012np} for two holomorphic curves $C_1,C_2$ one can compute the $(C_1+C_2)^2=C_1.C_1+C_2.C_2+2C_1.C_2$ when their self intersections and   $C_1.C_2$ is known. For the special case of $C_1.C_1=-2$ and $C_2.C_2=-2$  and $C_1$ intersecting $C_2$ at a single point.i.e. $C_1.C_2=1$ we get $(C_1+C_2)^2=-2-2+2(1)=-2$. Generalizing this to a chain of $(-2)$ curves  $C_1,C_2,...,C_N$ in which $i$-th curve intersects only $(i-1)$-th and $(i+1)$-th curves.i.e. $C_i.C_i=-2,i=1,...,N$ and $C_i.C_{i-1}=1,C_i.C_{i+1}=1$ with all other intersections equal to zero it is easy to see that 
\bea
(C_1+C_2+...+C_N)^2=-2
\eea
It is interesting to speculate that this property of the $(-2)$ curves mimics the result  (\ref{selfsim}) with the universal factor $W(\tau,m,\epsilon_1,\epsilon_2)$ playing the role of the identity element.\\
As shown in the previous sections, for multiple M2-branes between consecutive M5-branes,  instead of the free energies it is the elliptic genera that carry the recursive structure.  It will be interesting to elaborate on this phenomena in the framework of F-theory. 

 \section{Domain wall degrees of freedom: coupled supersymmetric WZW models}\label{wzw}
 We begin this section by reviewing the supergroup   WZW models. 
 The supergroup WZW model \cite{Okazaki:2015fiq,Okazaki:2016pne} is described by the maps $f:\Sigma\to (super group)\mbox{SG}$  from a two dimensional Euclidean Riemann surface $\Sigma$ to the supergroup $SG$ and its dynamics is given by the action
\bea\label{eq:WZWaction}
&S[f]=-\frac{k}{8\pi}\int_{\Sigma}d^2x(f^{-1}\partial^{\alpha}f,f^{-1}\partial_{\alpha}f)-\nonumber\\&\frac{ik}{24\pi}d^3x\epsilon^{\mu\nu\lambda}(f^{-1}\partial_{\mu}f,[f^{-1}\partial_{\nu}\lambda,f^{-1}\partial_{\lambda}f])\nonumber\\
\eea
where $\mbox{M}$ is a three-manifold with $\Sigma$ as its boundary and $k\in\mathbb{Z}$ is the level. The symmetry group $SG(z)\times SG(\bar{z})$ that generates the left and right actions is defined by
\bea\label{eq:WZWtrans}
f(z,\bar{z})\to \Lambda(z)f(z,\bar{z})\bar{\Lambda}^{-1}(\bar{z})
\eea
where $\Lambda(z)$ and $\Lambda(\bar{z})$ denote the arbitrary SG-valued functions of the complex variables
$z$ and $\bar{z}$. Note that the equation (\ref{eq:WZWaction}) is invariant under the transformation given by (\ref{eq:WZWtrans}). The conserved currents for this symmetry are given by
\bea
J(z)=J^a(z)T_a=-k\partial_zf.f^{-1}
\eea
with the generators of the Lie superalgebra $\bf{sg}$ denoted by $T^a$. The OPE of the generators $J^a$ is given by
\bea
J^a(z)J^b(w)\sim \frac{k(T^a,T^b)}{(z-w)^2}+\frac{[T^a,T^b]_cJ^c(w)}{z-w}
\eea
As a consequence of the OPE we have the following commutation relations which define the affine Lie superalgebra $\hat{\bf{sg}}$ \cite{Okazaki:2016pne}
\bea
[J^a_n,J^n_m]=[T^a,T^b]_cJ^c_{n+m}+m(T^a,T^b)\delta_{n+m}k
\eea
On the boundary of an M2-brane the description by the ABJM model gives rise to the WZW model. The content of the ABJM  model can be described in terms of type IIB branes configurations. To this end, see  Fig.\ref{MTV2}, note that the T-duality operation on $D3$-branes  wrapped on a circle gives rise to D2-branes in type IIA. These D2-branes can be lifted to the M2 branes in M-theory. To get the required contents of Chern-Simons description of the ABJM theory, the D3-branes are arranged so as to intersect two NS5-branes along the circle. Moreover the $k$ D5-branes are added to this configuration  as summarised in the table given in fig(\ref{MTV2}).
\begin{figure}
\begin{tabular}{ |p{3cm}|p{4cm}|  }
 \hline
 \multicolumn{2}{|c|}{type IIB  space-time} \\
 \hline
 & $x^0\hspace{0.1cm} x^1 \hspace{0.1cm} x^2\hspace{0.1cm} x^3\hspace{0.1cm} x^4\hspace{0.1cm} x^5\hspace{0.1cm} x^6\hspace{0.1cm} x^7\hspace{0.1cm} x^8\hspace{0.1cm} x^9$\\
 \hline
 NS5   & $\times\hspace{0.1cm} \times \hspace{0.1cm} \times\hspace{0.1cm} \times\hspace{0.1cm} \times\hspace{0.1cm} \times\hspace{0.1cm} \hspace{0.1cm} \hspace{0.1cm} \hspace{0.1cm} \hspace{0.1cm} $ \\
 D5&   $\times\hspace{0.1cm} \times \hspace{0.1cm}\times \hspace{0.1cm}\times \hspace{0.1cm}\times \hspace{0.1cm} \hspace{0.1cm} \hspace{0.1cm} \hspace{0.1cm}\hspace{0.1cm}\hspace{0.1cm}\hspace{0.1cm}\hspace{0.1cm}\times $ \\
 D3$_+$ &$\times\hspace{0.1cm} \times \hspace{0.1cm}\times \hspace{0.1cm} \hspace{0.1cm} \hspace{0.1cm} \hspace{0.1cm} \hspace{0.1cm} \hspace{0.1cm} \hspace{0.1cm} +\hspace{0.1cm} $ \\
  D3$_-$ &$\times\hspace{0.1cm} \times \hspace{0.1cm}\times \hspace{0.1cm} \hspace{0.1cm} \hspace{0.1cm} \hspace{0.1cm} \hspace{0.1cm} \hspace{0.1cm} \hspace{0.1cm} -\hspace{0.1cm} $ \\
 \hline
\end{tabular}
\caption{type IIB  picture}
 \label{MTV2}
\end{figure}
The $x^6$ direction is compact with period $2\pi R$, with the two NS5-branes located at $x^6=0$ and $x^6=\pi R $.  $x^6=0$ is the locus of D5-branes. Resolving the intersections of the NS5-brane with the $k$ D5-branes produces a $(p,q)$ 5-brane web. The resulting theory is  super Yang-Mills with massive chiral multiplets. Integrating out the chirals gives rise to the Chern-Simons theory. Finally the T-duality operation along $x^6$ followed by the lift to 11-dimensions gives rise to M2-branes spanning $(x^0,x^1,x^2)$. Under the T-duality the 5-branes turn into KK-monopoles and D6-branes. The low energy description is thus give by M2-branes probing $\mathbb{C}^4/\mathbb{Z}_k$. 

\begin{figure}
\begin{tabular}{ |p{3cm}|p{4cm}|  }
 \hline
 \multicolumn{2}{|c|}{11d M-theory space-time} \\
 \hline
 & $x^0\hspace{0.1cm} x^1 \hspace{0.1cm} x^2\hspace{0.1cm} x^3\hspace{0.1cm} x^4\hspace{0.1cm} x^5\hspace{0.1cm} x^6\hspace{0.1cm} x^7\hspace{0.1cm} x^8\hspace{0.1cm} x^9\hspace{0.1cm} x^{10}$\\
 \hline
 M5   & $\times\hspace{0.1cm} \times \hspace{0.1cm} \hspace{0.1cm} \hspace{0.1cm} \times\hspace{0.1cm} \times\hspace{0.1cm} \hspace{0.1cm} \hspace{0.1cm} \hspace{0.1cm} \hspace{0.1cm}\hspace{0.1cm}\hspace{0.1cm}\hspace{0.1cm}\hspace{0.1cm}\times\hspace{0.1cm}\times $ \\
 $M5^{\prime}$   & $\times\hspace{0.1cm} \times \hspace{0.1cm} \hspace{0.1cm} \hspace{0.1cm} \hspace{0.1cm} \hspace{0.1cm}\hspace{0.1cm} \hspace{0.1cm} \times \hspace{0.14cm} \times\hspace{0.1cm}\hspace{0.1cm}\hspace{0.1cm}\hspace{0.1cm}\hspace{0.1cm}\times\hspace{0.1cm}\times $ \\ \hline
  $M2$   & $\times\hspace{0.1cm} \times \hspace{0.1cm} \times \hspace{0.1cm} \hspace{0.1cm} \hspace{0.1cm}\hspace{0.1cm} \hspace{0.1cm}  \hspace{0.1cm} \hspace{0.1cm}\hspace{0.1cm}\hspace{0.1cm}\hspace{0.1cm}\hspace{0.1cm}\hspace{0.1cm} $ \\ \hline
\end{tabular}
\caption{the dual M-theory picture}
 \label{MTV3}
\end{figure}

In summary, the configuration of $\mbox{N}$ D3$^+$-branes and $\mbox{N}$ D3$^-$-branes that are stretched between $coincident$ NS5- and  NS5$^{\prime}$-branes can be lifted to the M-theory configuration M5-N M2-$M5^{\prime}$ to give a $\mbox{GL}(\mbox{N}|\mbox{N})$ WZW model. In the same way the configuration Fig.(\ref{MTV3}) M5-N M2-$M5^{\prime}$-M M2-M5 will give $\mbox{GL}(\mbox{N}|\mbox{N})\times \mbox{GL}(\mbox{M}|\mbox{M})$ WZW model with additional bi-fundamental matter content.\\
 \begin{figure}[h]

 \begin{tikzpicture}
\draw[black, thick] (2,4) -- (1,4)node[above]{$\nu_{i}$};

\draw[red, thick] (1.6,4.5) -- (2.4,4.5);

\draw[black, thick] (2,4) -- (2,5);
\draw[black, thick] (2,4) -- (2.8,3.3);

\draw[black, thick] (2.8,3.3) -- (3.8,3.3)node[below]{$\nu_{i+1}$};
\draw[black, thick] (2.8,3.3) -- (2.8,2.3);

\draw[red, thick] (2.4,2.8) -- (3.2,2.8);


\end{tikzpicture}
 \caption{toric diagram for  the open string wave function $\mathcal{W}_{\nu_i\nu_{i+1}}$. Red lines denote  the periodicity of the vertical side.}
\end{figure}
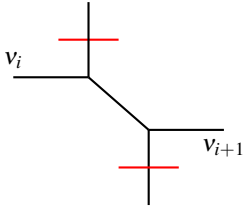
 
The open topological string wave function $\mathcal{W}_{\nu_i\nu_{i+1}}$ which is a building block of  the partition function (\ref{ZNTOP}),  is also the counting function for BPS excitations corresponding to the intersecting configurations of M2- and M5- branes. The topological open string wavefunction takes the form \cite{Haghighat:2013gba}
\bea\label{opentopw}
&\mathcal{W}_{\nu_i\nu_{i+1}}(Q_{\tau},Q_m,t,q)=t^{-\frac{||\nu_{m+1}||^2}{2}}q^{-\frac{||\nu_{m}||^2}{2}}
\tilde{Z}_{\nu^t_m}(q^{-1},t^{-1})\nonumber\\&\tilde{Z}_{\nu_{m+1}}(t^{-1},q^{-1})Q_m^{-\frac{|\nu_m|+|\nu_{m+1}|}{2}}\prod_{k=1}(1-Q_{\tau}^k)^{-1}\nonumber\\&\times\prod_{i,j=1}\frac{(1-Q_{\tau}^kQ_m^{-1}q^{\nu_{m+1,i}-j+\frac{1}{2}}t^{\nu_{m,j}-i+\frac{1}{2}})(1-Q_{\tau}^{k-1}Q_m q^{\nu_{m,i}-j+\frac{1}{2}}t^{\nu_{m+1,j}^t-i+\frac{1}{2}})}{(1-Q_{\tau}^k q^{\nu_{m+1,i}-j+1}t^{\nu_{m+1,j}-i})(1-Q_{\tau}^{k} q^{\nu_{m,i}-j}t^{\nu_{m,j}^t-i+1})}\nonumber\\
\eea
We can rewrite the factor $\bf{W}_{k_1k_2...k_r}$, given in eq. (\ref{eq:genconf}),  in terms of the open topological string wave function $\mathcal{W}_{\nu_a\nu_{a+1}}$.
Using the definition of $Z_{k_1k_2...k_r}$ \cite{Hohenegger:2015cba} as
\bea
Z_{k_1k_2...k_r}=(-1)^{k_1+k_2+k_3+...k_{n}}\sum_{\nu_a,|\nu_a|=k_a}\mathcal{W}_{\emptyset\nu_1}\mathcal{W}_{\nu_1\nu_2}\mathcal{W}_{\nu_2\nu_3}...\mathcal{W}_{\nu_n\emptyset}
\eea
we can write (\ref{eq:generalrecursive}) in terms of the topological string wavefunction $\mathcal{W}_{\nu\mu}$ as
\bea
&Z_{k_1k_2...k_r\emptyset \emptyset...\emptyset\emptyset...\emptyset}(\tau,m,\epsilon_1,\epsilon_2)\mbox{\bf{W}}_{k_1k_2...k_r}^2(\tau,m,\epsilon_1,\epsilon_2)=\nonumber\\
&(-1)^{k_1+k_2+k_3+...k_{3r}}\sum_{\nu_a,|\nu_a|=k_a}\mathcal{W}_{\emptyset\nu_1}\mathcal{W}_{\nu_1\nu_2}\mathcal{W}_{\nu_2\nu_3}...\mathcal{W}_{\nu_{r-1}\nu_r}\nonumber\\&\mathcal{W}_{\nu_{r+1}\nu_{r+2}}...\mathcal{W}_{\nu_{2r-1}\nu_{2r}}  \mathcal{W}_{\nu_{2r+1}\nu_{2r+2}}...\mathcal{W}_{\nu_{3r-1}\nu_{3r}}\mathcal{W}_{\nu_{3r}\emptyset} \nonumber\\&
\bigg(\mathcal{W}_{\nu_{r}\nu_{r+1}}-\mathcal{W}_{\nu_{r}\emptyset}\mathcal{W}_{\emptyset\nu_{r+1}} \bigg)\bigg(\mathcal{W}_{\nu_{2r}\nu_{2r+1}}-\mathcal{W}_{\nu_{2r}\emptyset}\mathcal{W}_{\emptyset\nu_{2r+1}}\bigg)
\eea

After normalising by the closed topological string partition function $\mathcal{W}_{\emptyset\emptyset}(Q_{\tau},Q,t,q)$ the resulting expression is
\bea\label{eq:Dmunu}
&D_{\nu_i\nu_{i+1}}(Q_{\tau},Q,t,q)=\frac{\mathcal{W}_{\nu_i\nu_{i+1}}(Q_{\tau},Q,t,q)}{\mathcal{W}_{\emptyset\emptyset}(Q_{\tau},Q,t,q)}=t^{-\frac{||\nu_{m+1}||^2}{2}}q^{-\frac{||\nu_{m}||^2}{2}}\nonumber\\&Q_m^{-\frac{|\nu_m|+|\nu_{m+1}|}{2}}\prod_{k=1}\prod_{(i,j)\in\nu_m}\nonumber\\&\times\frac{(1-Q_{\tau}^kQ_m^{-1}q^{-\nu_{m,i}+j-\frac{1}{2}}t^{-\nu_{m+1,j}+i-\frac{1}{2}})(1-Q_{\tau}^{k-1}Q_m q^{\nu_{m,i}-j+\frac{1}{2}}t^{\nu_{m+1,j}^t-i+\frac{1}{2}})}{(1-Q_{\tau}^k q^{\nu_{m,i}-j}t^{\nu_{m,j}^t-i+1})(1-Q_{\tau}^{k-1} q^{-\nu_{m,i}+j-1}t^{-\nu_{m,j}^t+i})}\prod_{(i,j)\in\nu_{m+1}}\nonumber\\&\times
\frac{(1-Q_{\tau}^kQ_m^{-1}q^{\nu_{m+1,i}-j+\frac{1}{2}}t^{\nu_{m,j}^t-i+\frac{1}{2}})(1-Q_{\tau}^{k-1}Q_m q^{-\nu_{m+1,i}+j-\frac{1}{2}}t^{-\nu_{m,j}^t+i-\frac{1}{2}})}{(1-Q_{\tau}^k q^{\nu_{m+1,i}-j+1}t^{\nu_{m+1,j}^t-i})(1-Q_{\tau}^{k-1} q^{-\nu_{m+1,i}+j}t^{-\nu_{m+1,j}^t+i-1})}\nonumber\\
&=
\Bigg(t^{-\frac{||\nu_{m+1}||^2}{2}}Q_m^{-\frac{|\nu_{m+1}|}{2}}\frac{(1-Q_m q^{\nu_{m,i}-j+\frac{1}{2}}t^{\nu_{m+1,j}^t-i+\frac{1}{2}})}{(1- q^{-\nu_{m,i}+j-1}t^{-\nu_{m,j}^t+i})}\nonumber\\&\times\prod_{k=1}\prod_{(i,j)\in\nu_m}\nonumber\\&
\frac{(1-Q_{\tau}^kQ_m^{-1}q^{-\nu_{m,i}+j-\frac{1}{2}}t^{-\nu_{m+1,j}+i-\frac{1}{2}})(1-Q_{\tau}^{k}Q_m q^{\nu_{m,i}-j+\frac{1}{2}}t^{\nu_{m+1,j}^t-i+\frac{1}{2}})}{(1-Q_{\tau}^k q^{\nu_{m,i}-j}t^{\nu_{m,j}^t-i+1})(1-Q_{\tau}^{k} q^{-\nu_{m,i}+j-1}t^{-\nu_{m,j}^t+i})}\Bigg)\nonumber\\&\times
\Bigg(q^{-\frac{||\nu_{m}||^2}{2}}Q_m^{-\frac{|\nu_m|}{2}}\frac{(1-Q_m q^{-\nu_{m+1,i}+j-\frac{1}{2}}t^{-\nu_{m,j}^t+i-\frac{1}{2}})}{(1- q^{-\nu_{m+1,i}+j}t^{-\nu_{m+1,j}^t+i-1})}\prod_{k=1}\nonumber\\&\times \prod_{(i,j)\in\nu_{m+1}}\nonumber\\&
\frac{(1-Q_{\tau}^kQ_m^{-1}q^{\nu_{m+1,i}-j+\frac{1}{2}}t^{\nu_{m,j}^t-i+\frac{1}{2}})(1-Q_{\tau}^{k}Q_m q^{-\nu_{m+1,i}+j-\frac{1}{2}}t^{-\nu_{m,j}^t+i-\frac{1}{2}})}{(1-Q_{\tau}^k q^{\nu_{m+1,i}-j+1}t^{\nu_{m+1,j}^t-i})(1-Q_{\tau}^{k} q^{-\nu_{m+1,i}+j}t^{-\nu_{m+1,j}^t+i-1})}\Bigg)\nonumber\\
\eea
 It can be interpreted \cite{Witten:1986bf} as the degrees of freedom of  two interacting WZW models .
To see this consider the first factor of eq.(\ref{eq:Dmunu}) 
\bea\label{minmodel}
&\Bigg(t^{-\frac{||\nu_{m+1}||^2}{2}}Q_m^{-\frac{|\nu_{m+1}|}{2}}\frac{(1-Q_m q^{\nu_{m,i}-j+\frac{1}{2}}t^{\nu_{m+1,j}^t-i+\frac{1}{2}})}{(1- q^{-\nu_{m,i}+j-1}t^{-\nu_{m,j}^t+i})}\prod_{k=1}\prod_{(i,j)\in\nu_m}\nonumber\\&
\frac{(1-Q_{\tau}^kQ_m^{-1}q^{-\nu_{m,i}+j-\frac{1}{2}}t^{-\nu_{m+1,j}+i-\frac{1}{2}})(1-Q_{\tau}^{k}Q_m q^{\nu_{m,i}-j+\frac{1}{2}}t^{\nu_{m+1,j}^t-i+\frac{1}{2}})}{(1-Q_{\tau}^k q^{\nu_{m,i}-j}t^{\nu_{m,j}^t-i+1})(1-Q_{\tau}^{k} q^{-\nu_{m,i}+j-1}t^{-\nu_{m,j}^t+i})}\Bigg)\nonumber\\
\eea
This factor is identical to the contribution of the field content of a supersymmetric WZW model \cite{Witten:1993jg,Berglund:1993fj,Henningson:1993nr}.

The expression (\ref{minmodel}) is reminiscent of the $\mathcal{N}=2$  minimal model product representation of the elliptic genus \cite{Witten:1993jg}.
It was suggested and proved in \cite{Berglund:1993fj,Henningson:1993nr,Nemeschansky:1994ac} that the minimal models  are equivalent to super-renormalizable Landau-Ginzburg models in the sense that the latter flows to the former at UV conformal point. To this end, use was made of the known characters of the discrete series representation of the $N=2$ superconformal algebra to compute its elliptic genus. For Landau-Ginzburg model a certain  superpotential deformation was used to render the exact computation of the elliptic genus possible.\\ It is desirable to independently compute the elliptic genus of the minimal model using some Lagrangian formulation of it.  It turned out certain minimal models have  Lagrangian description as  supersymmetric gauged WZW models.  The equivalence was shown by demonstrating  that the elliptic  genus computed for a Landau-Ginzburg model matches with the elliptic genus computed for a particular supersymmetric gauged WZW model  \cite{Henningson:1993nr}.\\
The field content of the supersymmetric gauged WZW model comprise  \cite{Henningson:1993nr} a Lie group G valued bosonic field $g$, a gauge field $A_{\alpha}$ that is $\mbox{Lie(H)}$-valued,where $\mbox{H}\subset \mbox{G}$ and $\mbox{Lie(G/H)}$ valued left moving and right moving fermionic fields $\widetilde{\psi}_+$ and $\widetilde{\psi}_-$ respectively. The dynamics is given by the Lagrangian
\bea
S&=&-\frac{k}{8\pi}\int d^2z\sqrt{h}h^{ij}Trg^{-1}\partial_ig.g^{-1}\partial_jg\nonumber\\&-&\frac{ik}{12\pi}\int_Bd^2\sigma\epsilon^{ijk}Trg^{-1}\partial_ig.g^{-1}\partial_jg.g^{-1}\partial_kg\nonumber\\&+&\frac{k}{2\pi}\int d^2z Tr(A_{\bar{z}}g^{-1}\partial_zg-A_z\partial_{\bar{z}}gg^{-1}-A_{\bar{z}}A_z+A_{\bar{z}}g^{-1}A_zg)\nonumber\\&+&\frac{ik}{4\pi}\int d^2z Tr(\widetilde{\psi}_+D_{\bar{z}}\widetilde{ \psi}_++\widetilde{\psi}_-D_z\widetilde{\psi}_-)\nonumber\\
\eea
where $\mbox{B}$ is the 3-manifold whose boundary is the 2d worldsheet, $h_{ij}$ is the worldsheet metric, $D_z=\partial_z+[A_z,],D_{\bar{z}}=\partial_{\bar{z}}+[A_{\bar{z}},]$ are the covariant derivatives and the integer k is the level.
After the identification of a global $U(1)$ that is part of the left moving $N=2$ algebra, the charge assignment of the fields is given by
\bea
\widetilde{\psi}_+&\to& e^{i\frac{\gamma}{(k+2)}}\widetilde{\psi}_+\nonumber\\
\widetilde{\psi}_-&\to& e^{i\frac{\gamma(k+1)}{(k+2)}}\widetilde{\psi}_-\nonumber\\
g&\to&-i\frac{\gamma}{(k+2)}(Ug+gU)\nonumber\\
A_{\alpha}&\to& A_{\alpha}
\eea
where $\mbox{U}\in \mbox{Lie SU(2)}$ denotes the generator of the $\mbox{U(1)}\subset \mbox{SU(2)}$ which is gauged.
This allows to show that the elliptic genus of  the supersymmetric WZW model of the coset $\mbox{SU(2)/U(1)}$ is given by 
\bea\label{eq:1}
\mathcal{E}^{SWZW}&=&e^{-i\gamma k\alpha/2}\frac{1-e^{i\gamma (k+1)\alpha}}{1-e^{i\gamma\alpha}}\nonumber\\&\times&\prod_{n=1}^{\infty}\frac{(1-q^ne^{i\gamma (k+1)\alpha})(1-q^ne^{-i\gamma (k+1)\alpha})}{(1-e^{i\gamma\alpha})(1-e^{-i\gamma\alpha})}\nonumber\\
\eea
where the contribution of the fermonic  zero modes is separated respectively as
\bea
\frac{1}{(1-e^{i\gamma \alpha})(1-e^{-i\gamma\alpha})}
\eea
and bosonic zero modes as
\bea
e^{-i\gamma k\alpha/2}(1-e^{i\gamma (k+1)\alpha})(1-e^{-i\gamma\alpha})
\eea
Moreover the non-zero modes contribution of the fermions as
\bea
(1-q^ne^{i\gamma (k+1)\alpha})(1-q^ne^{-i\gamma (k+1)\alpha})(1-\bar{q}^ne^{i\gamma \alpha})(1-\bar{q}^ne^{-i\gamma \alpha})
\eea
and the bosons as
\bea
\frac{1}{(1-q^ne^{i\gamma \alpha})(1-q^ne^{-i\gamma \alpha})(1-\bar{q}^ne^{i\gamma \alpha})(1-\bar{q}^ne^{-i\gamma \alpha})}
\eea
Note that the anti-holomorphic part  is cancelled out.\\

The various contributions of supersymmetric WZW model are thus arranged in the following:
\subsection*{$\bullet$\bf{ contribution of the fermionic and  bosonic zero modes}:}
\bea
\frac{(1-Q_m q^{\nu_{m,i}-j+\frac{1}{2}}t^{\nu_{m+1,j}^t-i+\frac{1}{2}})}{(1- q^{-\nu_{m,i}+j-1}t^{-\nu_{m,j}^t+i})}
\eea
\subsection*{$\bullet$ \bf{contribution of the fermionic and bosonic non-zero modes}:}
\bea
&\bigg(
\prod_{k=1}\prod_{(i,j)\in\nu_m}\nonumber\\&\frac{(1-Q_{\tau}^kQ_m^{-1}q^{-\nu_{m,i}+j-\frac{1}{2}}t^{-\nu_{m+1,j}+i-\frac{1}{2}})(1-Q_{\tau}^{k}Q_m q^{\nu_{m,i}-j+\frac{1}{2}}t^{\nu_{m+1,j}^t-i+\frac{1}{2}})}{(1-Q_{\tau}^k q^{\nu_{m,i}-j}t^{\nu_{m,j}^t-i+1})(1-Q_{\tau}^{k} q^{-\nu_{m,i}+j-1}t^{-\nu_{m,j}^t+i})}\nonumber\\&
\eea
\subsection*{$\bullet$ \bf{contribution from  the phase factors}:}
\bea
&\bigg(t^{-\frac{||\nu_{m+1}||^2}{2}}Q_m^{-\frac{|\nu_{m+1}|}{2}}\bigg)
\eea
Note that for the second WZW model the phase contribution  changes from  $t^{-\frac{||\nu_{m+1}||^2}{2}}$ to $q^{-\frac{||\nu_{m}||^2}{2}}$ i.e. from a $t$-factor to a $q$-factor. Similarly the second factor in eq.(\ref{eq:Dmunu})  also describes a WZW model. 
\\
Recall that the partition function $Z_{N}(\tau,m,t_{f_1},t_{f_2},...,t_{f_N},\epsilon_1,\epsilon_2)$ of $N$ parallel and separated M5-branes with M2-branes stretched between them can alternatively\cite{Haghighat:2013gba}  be written in terms of the normalised open topological  string wavefunctions $D_{\nu_i\nu_{i+1}}(Q_{\tau},Q,t,q)$ as
\bea
&Z_{N}(\tau,m,t_{f_1},t_{f_2},...,t_{f_N},\epsilon_1,\epsilon_2)=\nonumber\\&\sum_{\nu_1,...,\nu_{N-1}}(\prod_{a=1}^{N-1}(-Q_{f_a})^{|\nu_a|})\times D_{\emptyset\nu_1}(Q_{\tau},Q,t,q)\nonumber\\&\times D_{\nu_1\nu_2}(Q_{\tau},Q,t^{-1},q^{-1})D_{\nu_2\nu_3}(Q_{\tau},Q,t,q)...D_{\nu_{N-1}\emptyset}
\eea
This form of the partition function allows an interpretation in terms of $N$ domain walls interpolating between the M2-branes vacua. In terms of the supersymmetric WZW model, we can say that the partition function is a superposition of the wavefunctions of a chain of coupled supersymmetric WZW models.  
The centre of mass motion of the multiple M-strings as well as their mutual dynamics is encoded in this wave function. For example for the case of two M-strings it involves their centre of mass motion as well as their motion relative to each other. The components of the elliptic genus $Z_{\nu_1\nu_2,...,\nu_k}$ are related to the open topological string  wave function. For example
\bea
Z_{22}-Z_{2\emptyset}^2&=&\mathcal{W}_{\emptyset 2}\mathcal{W}_{2\emptyset}\bigg(
\mathcal{W}_{22}-\mathcal{W}_{\emptyset 2}\mathcal{W}_{2\emptyset }\bigg)\nonumber\\
Z_{33}-Z_{3\emptyset}^2&=&\mathcal{W}_{\emptyset 3}\mathcal{W}_{3\emptyset}\bigg(
\mathcal{W}_{33}-\mathcal{W}_{\emptyset 3}\mathcal{W}_{3\emptyset }\bigg)\nonumber\\
Z_{1212}-Z_{12\emptyset\emptyset}^2&=&\mathcal{W}_{\emptyset 1}\mathcal{W}_{12}\mathcal{W}_{2\emptyset}\bigg(\mathcal{W}_{21}\mathcal{W}_{12}-\mathcal{W}_{\emptyset1}\mathcal{W}_{12} W_{2\emptyset}  \bigg)
\eea
In other words the universal factors $W_{2}(\tau,m,\epsilon_1,\epsilon_2),W_{3}(\tau,m,\epsilon_1,\epsilon_2$ and $W_{12}(\tau,m,\epsilon_1,\epsilon_2)$ for these M5-M2 branes configurations can be expressed in terms of open topological wavefunction in eq.(\ref{opentopw}) as
\bea
W_{2}(\tau,m,\epsilon_1,\epsilon_2)&=&\bigg(
\mathcal{W}_{22}-\mathcal{W}_{\emptyset 2}\mathcal{W}_{2\emptyset }\bigg)\nonumber\\
W_{3}(\tau,m,\epsilon_1,\epsilon_2)&=&\bigg(
\mathcal{W}_{33}-\mathcal{W}_{\emptyset 3}\mathcal{W}_{3\emptyset }\bigg)\nonumber\\
W_{12}(\tau,m,\epsilon_1,\epsilon_2)&=&\bigg(\mathcal{W}_{21}\mathcal{W}_{12}-\mathcal{W}_{\emptyset1}\mathcal{W}_{12} W_{2\emptyset}  \bigg)
\eea
Similarly we can write for   $W_{N}(\tau,m,\epsilon_1,\epsilon_2)$
\bea
W_{N}(\tau,m,\epsilon_1,\epsilon_2)&=&\bigg(
\mathcal{W}_{NN}-\mathcal{W}_{\emptyset N}\mathcal{W}_{N\emptyset }\bigg)\nonumber\\
\eea
 Recall that in a given M-theory vacuum the coupling constant $\tau$ is related to the radius of the circle $S^1$ parallel to the   M5-brane worldvolume.
Formally we can consider different coupling constants $\tau_i$ for different domain walls.
Each $\tau_i$ is related to the circle $S^1$ parallel to the  $i$-th M5-brane worldvolume.This M-theory set up can be dualized in type IIB strings to a 5d $\mathcal{N}=1^*$ supersymmetric gauge theory living on a particular $(p,q)$ D5-NS5-branes web. The $\tau_i$ correspond to the gauge coupling constant of the supersymmetric gauge theories dual to corresponding M5 brane-M2 brane-Mstring configurations. For these general cases for instance we can write
\bea
&Z_{1212}-Z_{12\emptyset\emptyset}Z_{\emptyset12\emptyset}
 \nonumber\\
&= \mathcal{W}_{\emptyset 1}(\tau_1)\mathcal{W}_{12}(\tau_4)\mathcal{W}_{2\emptyset}(\tau_5)\bigg( \mathcal{W}_{12}(\tau_2)\mathcal{W}_{21}(\tau_3)-
\nonumber\\&\mathcal{W}_{\emptyset 1}(\tau_3)\mathcal{W}_{12}(\tau_2)\mathcal{W}_{2\emptyset}(\tau_3)\bigg)\nonumber\\
\eea
\bea
&Z_{222}-Z_{22\emptyset}Z_{\emptyset\emptyset 2}-Z_{\emptyset22}Z_{2\emptyset\emptyset}-Z_{2\emptyset\emptyset}Z_{\emptyset2\emptyset}Z_{\emptyset\emptyset2}\nonumber\\&
= \mathcal{W}_{\emptyset 2}(\tau_1)\mathcal{W}_{2\emptyset}(\tau_4)\bigg( (\mathcal{W}_{22}(\tau_2)-\mathcal{W}_{\emptyset2}(\tau_2)\mathcal{W}_{2\emptyset}(\tau_2))\nonumber\\&\times(\mathcal{W}_{22}(\tau_3)-\mathcal{W}_{\emptyset2}(\tau_3)\mathcal{W}_{2\emptyset}(\tau_3))\bigg)\nonumber\\
\eea
 More generally  we can write for $\nu_1=\nu_2=\nu_3$ the recursive relation for $Z_{\nu_1\nu_2\nu_2}$ as
 \bea
 Z_{\nu_1\nu_2\nu_3}&-&Z_{\nu_1\nu_2}Z_{\nu_3}-Z_{\nu_1}Z_{\nu_2\nu_3}+Z_{\nu_1}Z_{\nu_2}Z_{\nu_3}\nonumber\\&=&
 \mathcal{W}_{\emptyset\nu_1}\mathcal{W}_{\emptyset\nu_3}\big(\mathcal{W}_{\nu_1\nu_2} -\mathcal{W}_{\emptyset\nu_1} \mathcal{W}_{\emptyset\nu_2}  \big)\big(\mathcal{W}_{\nu_2\nu_3} -\mathcal{W}_{\emptyset\nu_2} \mathcal{W}_{\emptyset\nu_3}  \big)\nonumber\\
 \eea
 Comparing the last expression with equation (\ref{eq:fusion}), we see that the first factor $\mathcal{W}_{\emptyset\nu_1}\mathcal{W}_{\emptyset\nu_3}$ is the result of fusing all the partitions, the second factor $\big(\mathcal{W}_{\nu_1\nu_2} -\mathcal{W}_{\emptyset\nu_1} \mathcal{W}_{\emptyset\nu_2}  \big)$ appears when we fuse the partitions $\nu_1,\nu_2$ along with the removal of the second M5-brane and the third factor $\big(\mathcal{W}_{\nu_2\nu_3} -\mathcal{W}_{\emptyset\nu_2} \mathcal{W}_{\emptyset\nu_3}  \big)$ appears when we remove the third M5-brane fusing $\nu_2,\nu_3$.\\
 A generalization of  $Z_{\nu_1\nu_2\nu_3\nu_4...\nu_k}$  for  $\nu_1=\nu_2=\nu_3=...=\nu_k$ can be expressed as
 \bea
 &Z_{\nu_1\nu_2\nu_3\nu_4...\nu_k}-Z_{\nu_1\nu_2\nu_3...\nu_{k-1}}Z_{\nu_k}-Z_{\nu_2\nu_3\nu_4...\nu_{k}}Z_{\nu_1}+Z_{\nu_1}Z_{\nu_2}Z_{\nu_3\nu_4...\nu_{k}}\nonumber\\&+Z_{\nu_{k-1}}Z_{\nu_k}Z_{\nu_1\nu_2\nu_3\nu_4...\nu_{k-2}}+...-Z_{\nu_1}Z_{\nu_2}Z_{\nu_3}...Z_{\nu_k}\nonumber\\&=
 \mathcal{W}_{\emptyset\nu_1}\mathcal{W}_{\emptyset\nu_4}\big(\mathcal{W}_{\nu_1\nu_2} -\mathcal{W}_{\emptyset\nu_1} \mathcal{W}_{\emptyset\nu_2}  \big)\big(\mathcal{W}_{\nu_2\nu_3} -\mathcal{W}_{\emptyset\nu_2} \mathcal{W}_{\emptyset\nu_3}  \big)\nonumber\\
&\times\big(\mathcal{W}_{\nu_3\nu_4} -\mathcal{W}_{\emptyset\nu_3} \mathcal{W}_{\emptyset\nu_4}  \big) ...\big(\mathcal{W}_{\nu_{k-1}\nu_k} -\mathcal{W}_{\emptyset\nu_{k-1}} \mathcal{W}_{\emptyset\nu_k}  \big)
 \eea

\section{ABJM model vs M-theory}\label{MABJM}
The ABJM model is defined by a 3d $\mathcal{N}=6$ supersymmetric $U(N)_k\times U(N)_{-k}$ Chern-Simons theory with matter coupling given by the bifundamental scalars $Z_a$ and spinors $\Psi^a$ with $SU(4)$ R-symmetry index $a$.
The  low energy 2d gauge theory  corresponds to the reduction of the worldvolume theory of M2-branes to two dimensions with the boundary conditions provided by the M5-branes. For details we refer the reader to \cite{Hosomichi:2014rqa}.  This 2d theory is termed as ABJM $slab$  and is identical to the $\mathcal{N}=(4,4)$ super Yang-Mills having $SU(2)^3$ R-symmetry and the gauge coupling $g^2_{2d}$ is determined by the distance between the M2-branes stack and the $\mathbb{Z}_k$ orbifold singularity of the transverse space $\mathbb{C}^2/\mathbb{Z}_k$. This 2d theory is special in the sense that the M2-branes do not sit on top of the $\mathbb{Z}_k$ singularity. This avoids the appearance  of Nahm poles. Moreover the Ramond-Ramond boundary conditions used in the definition of elliptic genus project out the massive modes corresponding to the KK modes.  
The elliptic genus of the 2d gauge theory with non-zero coupling constant $g^2_{2d}>0$ turns out to be the same as that of $\mathcal{N}=(4,4)$ super Yang Mills.\\
The M2-M5 branes intersection is described by the boundary conditions that preserve six supercharges and the $SU(2)\times SU(2)\times U(1)$ subgroup of the full R-symmetry group $SU(4)$. The two scalars $Z_1,Z_2$ are longitudinal to the M5-brane and form a doublet under one of the two $SU(2)s$. The other two scalars $Z_3,Z_4$ are transverse and form a doublet under the second $SU(2)$. Moreover under the $U(1)$ group the two doublets are oppositely charged. The boundary conditions on bosons and fermions are given as
\bea
&&\Psi_+^1=\Psi_+^2=\Psi_-^1=\Psi_-^2=0,\quad \bar{\Psi}_+^1=\bar{\Psi}_+^2=\bar{\Psi}_-^1=\bar{\Psi}_-^2=0,\nonumber\\ &&
D_{\mu}Z_A=0,\quad D_{x_2}Z_I=\frac{2\pi}{k}(Z_I\bar{Z}^JZ_J-Z_J\bar{Z}^JZ_I)\nonumber\\&&
Z_A\bar{Z}^IZ_B=Z_B\bar{Z}^IZ_A,\quad Z_I\bar{Z}^AZ_B=Z_B\bar{Z}^AZ_I
\eea
The boundary forces the gauge fields of the two $U(N)s$ to be related as
\bea
F_{\mu\nu}Z_A=Z_A\tilde{F}_{\mu\nu}
\eea
Moreover the variation of the ABJM action gives rise to a boundary term 
\bea\label{eq:boundaryS}
\delta S=\frac{k}{4\pi}\int_{boundary}Tr(\alpha dA-\tilde{\alpha}d\tilde{A})
\eea
which vanishes only if $A=\tilde{A}$ and $\alpha=\tilde{\alpha}$. However if one takes $A\ne \tilde{A}$ then the anomalous boundary term can be cancelled by introducing boundary fermions of one chirality coupled to $A$ and boundary fermions of the opposite chirality coupled to $\tilde{A}$. This effectively gives rise to WZW model degrees of freedom at the boundary and the gauge anomaly they generate cancels the anomalous term (\ref{eq:boundaryS}).\\
We first write down the expression of the elliptic genus of the 2d gauge theory obtained from the dimensional reduction  of the ABJM model for the case $k=1$ as considered in \cite{Hosomichi:2014rqa}
\bea\label{eq:ABJMint}
Z^{ABJM}_{\mbox{T}^2}=\int\prod_{i=1}^N \frac{dw_i d\bar{w}_i}{Im\tau}\prod_{i,j}\frac{\theta_1(w_i-w_j+m+\epsilon_+){\theta_1(w_i-w_j+m-\epsilon_+)}}{{\theta_1(w_i-w_j+\epsilon_1)\theta_1(w_i-w_j+\epsilon_2)}}
\eea

However in our case 
 the integral is finite with respect to the integration variables and no special regularization is required. 
 We will use instead a prescription given in \cite{Gadde:2013dda} for the case of $N=2$.

For  ABJM theory and   for $N=1$ we get the expression
\bea
Z^{ABJM}_{\mbox{T}^2}=\frac{\theta_1(m+\epsilon_+){\theta_1(m-\epsilon_+)}}{{\theta_1(\epsilon_1)\theta_1(\epsilon_2)}}
\eea
This expression matches with its  $Z^{M-string}_{\mbox{T}^2}$ \cite{Haghighat:2013gba}.\\
\subsection*{\underline{N$\ge$ 2}}
For the complex integration  beyond and including $N=2$ and we will use the results  \cite{Gadde:2013dda,Gadde:2015tra,Benini:2018hjy}
\bea
&Z_{\mbox{T}^2}^{ABJM}=\bigg(\frac{\theta_1(m+\epsilon_+){\theta_1(m-\epsilon_+)}}{{\theta_1(\epsilon_1)\theta_1(\epsilon_2)}}\bigg)^N\nonumber\\&\int\prod_{i=1}^N \frac{dw_i d\bar{w}_i}{Im\tau}\prod_{j\ne i}\frac{\theta_1(w_i-w_j+m+\epsilon_+){\theta_1(w_i-w_j+m-\epsilon_+)}}{{\theta_1(w_i-w_j+\epsilon_1)\theta_1(w_i-w_j+\epsilon_2)}}\nonumber\\
\eea
The $w_i,\bar{w}_i$ integrals are well defined and do not require any regularisation. So we can write normalised elliptic genus in the limit $Q_m=\sqrt{\frac{t}{q}}$ as
\bea\label{eq:ABJMN}
\widehat{Z}_{\mbox{T}^2}^{ABJM}&=&\int\prod_{i=1}^N \frac{dz_i }{2\pi i z_i}\prod_{j\ne i}\frac{\theta_1(\frac{z_i}{z_j} \frac{t}{q})\theta_1(\frac{z_i}{z_j})}{\theta_1(\frac{z_i}{z_j}t)\theta_1(\frac{z_i}{z_j}q^{-1})}\nonumber\\
\eea
For 2d gauge theories containing the adjoint matter the  poles contributing to the elliptic genus were found to be  \cite{Gadde:2015tra,Benini:2018hjy}
\bea
z_a=t^{x}q^{-y}
\eea
where $(x,y)$ are the coordinates of the $a$-th box in the Young diagram $\mu$ such that $|\mu|=N$ for all $N\ge 0$. Evaluating the residue of (\ref{eq:ABJMN}) on these poles we get
\bea
&\widehat{Z}_{\mbox{T}^2}^{ABJM}=\sum\prod^{x1\ne x_2,y_1\ne y_2}_{(x_1,y_1)\in Y,(x_2,y_2)\in Y^t} \nonumber\\&\frac{\theta_1((x_1-x_2)\epsilon_1+(y_1-y_2)\epsilon_2)\theta_1((x_1-x_2-1)\epsilon_1+(y_1-y_2-1)\epsilon_2)}{\theta_1((x_1-x_2-1)\epsilon_1+(y_1-y_2)\epsilon_2)\theta_1((x_1-x_2)\epsilon_1+(y_1-y_2+1)\epsilon_2)}\nonumber\\
&=\prod \frac{\theta_1(\epsilon_1)\theta_1(\epsilon_2)}{\theta_1^{\prime}\theta_1(\epsilon_1+\epsilon_2)} \sum\prod_{(x_1,y_1)\in Y,(x_2,y_2)\in Y^t} \nonumber\\&\times\frac{\theta_1((x_1-x_2)\epsilon_1+(y_1-y_2)\epsilon_2)\theta_1((x_1-x_2-1)\epsilon_1+(y_1-y_2-1)\epsilon_2)}{\theta_1((x_1-x_2-1)\epsilon_1+(y_1-y_2)\epsilon_2)\theta_1((x_1-x_2)\epsilon_1+(y_1-y_2+1)\epsilon_2)}\nonumber\\
\eea
This can be written in the canonical form by using the theorem (2.11) of reference \cite{Nakajima:2003pg}
\bea
&\sum_{(x_1,y_1)\in Y}t_1^{x_1}t_2^{y_1}+\sum_{(x_2,y_2)\in Y^t}t_1^{1-x_2}t_2^{1-y_2}-\nonumber\\&\bigg(\sum_{(x_1,y_1)\in Y}t_1^{x_1-\lambda^t_{y_1}(Y^t)}t_2^{1-y_1+\lambda_{x_1}(Y)}+\nonumber\\&\sum_{(x_2,y_2)\in Y^t}t_1^{1-x_2+\lambda^t_{y_2}(Y)}t_2^{y_2-\lambda_{x_2}(Y^t)}\bigg)=\nonumber\\
&\sum_{(x_1,y_1)\in Y,(x_2,y_2)\in Y^t}t_1^{x_1-x_2}t_2^{y_1-y_2}(1-t_1)(1-t_2)
\nonumber\\
\eea
as
\bea
&\widehat{Z}_{\mbox{T}^2}^{ABJM}=\prod \frac{\theta_1(\epsilon_1)\theta_1(\epsilon_2)}{\theta_1^{\prime}\theta_1(\epsilon_1+\epsilon_2)} \sum_{Y,Y^t}\prod_{(x_1,y_1)\in Y,(x_2,y_2)\in Y^t} \nonumber\\ &\times\frac{\theta_1(x_1\epsilon_1-y_1\epsilon_2)\theta_1((1-x_2)\epsilon_1-(1-y_2)\epsilon_2)}{\theta_1\big((x_1-\lambda_{y_1}^t)\epsilon_1-(1-y_1+\lambda_{x_1})\epsilon_2\big)\theta_1\big((1-x_2-\lambda_{y_2}^t)\epsilon_1-(y_2-\lambda_{x_2})\epsilon_2\big)}\nonumber\\
\eea
It is interesting to compare $\widehat{Z}_{\mbox{T}^2}$ with $Z^{IIA}_{\mbox{T}^2}$ given by
\bea
&Z^{IIA}_{\mbox{T}^2}=\sum_{Y,Y^t}\prod_{(i,j)\in Y}\nonumber\\ &\times\frac{\theta_1((i-1)\epsilon_1+(j-1)\epsilon_2)\theta_1(-i\epsilon_1-j\epsilon_2)}{\theta_1((i-\lambda^t_j)\epsilon_1+(\lambda_i-j+1)\epsilon_2)\theta_1((1-i+\lambda_j^t)\epsilon_1+(j-\lambda_i)\epsilon_2)}
\eea
\section{Conclusions}

We have studied the structure of the free energies of M-strings.
 An interesting recursive structure in the free energies (BPS counting functions) was observed \cite{Hohenegger:2015cba} for the configuration  ---M2-M5-M2-M5-M2--- of M2-M5 branes. We show that for configurations containing multiple M2-branes sandwiched between M5-branes the recursive structure in free energies is lost. Instead the coefficients $Z_{A_1A_2...A_n}$ in the expansion of partition function enjoy the recursive structure. 
For completeness we also describe the M2-branes configurations with symmetric representations and mixed representations. \\
The partition functions of M2-branes configuration that enter the M-strings elliptic genera can also be interpreted as the vacuum-n to vacuum-(n+1)  amplitude with M5-branes acting as the domain wall. The M5-brane domain wall act as the duality transformation that interpolates between the two vacua 
\cite{Gaiotto:2008ak}. ABJM formulation of M2-branes theories allows a more direct study of the domain walls partition functions. We compute the elliptic genus of a dimensionally reduced 2d theory of ABJM slab model and compare it with the M-string computations. We find an interesting mismatch that can be explained in terms of the centre of mass motion of M2-M5 branes in the transverse space. The  factor corresponding to the mismatch accounts for this centre of mass motion in the transverse space.\\
It will be interesting to study the WZW-topological string correspondence for more general backgrounds in M-theory.

\section*{Conflict of interest statement}
The author declares no conflicts of interest.
\section*{Acknowledgements}
\addcontentsline{toc}{section}{Acknowledgement}
The author would like to thank Amer Iqbal for giving useful   comments. The author is grateful to the referees for very useful comments. 
 Moreover the support provided by the Abdus Salam School of Mathematical Sciences, Lahore,  is gratefully  acknowledged.

\begin{appendices}
\section{Denominator factors\\ $D_{\nu_1,...,\nu_k}(\tau,m,\epsilon_1,\epsilon_2)$}\label{Den}
\bea
&D_{134134134}(\tau,m,\epsilon_1,\epsilon_2)= \theta _1\left(\epsilon _1-m\right)^9 \theta _1\left(-m-4 \epsilon _2\right)^3 \theta
   _1\left(-m-3 \epsilon _2\right)^6 \nonumber\\&\theta _1\left(-m+\epsilon _1-3 \epsilon _2\right)^3
   \theta _1\left(-m-2 \epsilon _2\right)^6\times \theta _1\left(-m+\epsilon _1-2 \epsilon
   _2\right)^6\nonumber\\&  \theta _1\left(-m-\epsilon _2\right)^9 \theta _1\left(-m+\epsilon _1-\epsilon
   _2\right)^6\nonumber
   \eea
   \bea
   &D_{134134\emptyset\emptyset\emptyset}(\tau,m,\epsilon_1,\epsilon_2)= \theta _1\left(\epsilon _1-m\right)^6 \theta _1\left(-m-4 \epsilon _2\right)^2 \theta
   _1\left(-m-3 \epsilon _2\right)^4\nonumber\\& \theta _1\left(-m+\epsilon _1-3 \epsilon _2\right)^2
   \theta _1\left(-m-2 \epsilon _2\right)^4 \times\theta _1\left(-m+\epsilon _1-2 \epsilon
   _2\right)^4\nonumber\\& \theta _1\left(-m-\epsilon _2\right)^6 \theta _1\left(-m+\epsilon _1-\epsilon
   _2\right)^4\nonumber
   \eea
   \bea
 &  D_{134\emptyset\emptyset\emptyset\emptyset\emptyset\emptyset}(\tau,m,\epsilon_1,\epsilon_2)= \theta _1\left(\epsilon _1-m\right)^3 \theta _1\left(-m-4 \epsilon _2\right) \theta
   _1\left(-m-3 \epsilon _2\right)^2\nonumber\\& \theta _1\left(-m+\epsilon _1-3 \epsilon _2\right)
   \theta _1\left(-m-2 \epsilon _2\right)^2\times \theta _1\left(-m+\epsilon _1-2 \epsilon
   _2\right)^2\nonumber\\& \theta _1\left(-m-\epsilon _2\right)^3 \theta _1\left(-m+\epsilon _1-\epsilon
   _2\right)^2\nonumber
   \eea
   \bea
 &  D_{232323}(\tau,m,\epsilon_1,\epsilon_2)= \theta _1\left(\epsilon _1-m\right)^6 \theta _1\left(-m-3 \epsilon _2\right)^3 \theta
   _1\left(-m-2 \epsilon _2\right)^6\nonumber\\& \theta _1\left(-m+\epsilon _1-2 \epsilon _2\right)^3\times 
   \theta _1\left(-m-\epsilon _2\right)^6\nonumber\\& \theta _1\left(-m+\epsilon _1-\epsilon _2\right)^6\nonumber
   \eea
   \bea
&   D_{2323\emptyset\emptyset}(\tau,m,\epsilon_1,\epsilon_2)= \theta _1\left(\epsilon _1-m\right)^4 \theta _1\left(-m-3 \epsilon _2\right)^2 \theta
   _1\left(-m-2 \epsilon _2\right)^4\nonumber\\& \theta _1\left(-m+\epsilon _1-2 \epsilon _2\right)^2\times 
   \theta _1\left(-m-\epsilon _2\right)^4\nonumber\\& \theta _1\left(-m+\epsilon _1-\epsilon _2\right)^4\nonumber
   \eea
   \bea
&   D_{232323}(\tau,m,\epsilon_1,\epsilon_2)= \theta _1\left(\epsilon _1-m\right)^2 \theta _1\left(-m-3 \epsilon _2\right) \theta
   _1\left(-m-2 \epsilon _2\right)^2\nonumber\\& \theta _1\left(-m+\epsilon _1-2 \epsilon _2\right)\times
   \theta _1\left(-m-\epsilon _2\right)^2\nonumber\\& \theta _1\left(-m+\epsilon _1-\epsilon _2\right)^2\nonumber
   \eea
   \bea
&   D_{121212}(\tau,m,\epsilon_1,\epsilon_2)= \theta _1\left(\epsilon _1-m\right)^6 \theta _1\left(-m-2 \epsilon _2\right)^3 \nonumber\\&\theta
   _1\left(-m-\epsilon _2\right)^6 \theta _1\left(-m+\epsilon _1-\epsilon _2\right)^3\nonumber
   \eea
   \bea
&   D_{1212\emptyset\emptyset}(\tau,m,\epsilon_1,\epsilon_2)= \theta _1\left(\epsilon _1-m\right)^4 \theta _1\left(-m-2 \epsilon _2\right)^2\nonumber\\& \theta
   _1\left(-m-\epsilon _2\right)^4 \theta _1\left(-m+\epsilon _1-\epsilon _2\right)^2
\nonumber
\eea
\bea
&   D_{12\emptyset\emptyset\emptyset\emptyset}(\tau,m,\epsilon_1,\epsilon_2)= \theta _1\left(\epsilon _1-m\right)^2 \theta _1\left(-m-2 \epsilon _2\right)\nonumber\\& \theta
   _1\left(-m-\epsilon _2\right)^2 \theta _1\left(-m+\epsilon _1-\epsilon _2\right)\nonumber
\eea

\end{appendices}
\nocite{<key>}
\bibliographystyle{utphys}

\bibliography{bibliography}

\providecommand{\href}[2]{#2}\begingroup\raggedright\begin{thebibliography}{10}

\bibitem{Haghighat:2013gba}
B.~Haghighat, A.~Iqbal, C.~Koz\c{c}az, G.~Lockhart, and C.~Vafa,
  ``{M-Strings},'' \href{http://dx.doi.org/10.1007/s00220-014-2139-1}{{\em
  Commun. Math. Phys.} {\bfseries 334} no.~2, (2015) 779--842},
  \href{http://arxiv.org/abs/1305.6322}{{\ttfamily arXiv:1305.6322 [hep-th]}}.

\bibitem{Haghighat:2013tka}
B.~Haghighat, C.~Kozcaz, G.~Lockhart, and C.~Vafa, ``{Orbifolds of
  M-strings},'' \href{http://dx.doi.org/10.1103/PhysRevD.89.046003}{{\em Phys.
  Rev. D} {\bfseries 89} no.~4, (2014) 046003},
  \href{http://arxiv.org/abs/1310.1185}{{\ttfamily arXiv:1310.1185 [hep-th]}}.

\bibitem{Hohenegger:2013ala}
S.~Hohenegger and A.~Iqbal, ``{M-strings, elliptic genera and $\mathcal{N} = 4$
  string amplitudes},'' \href{http://dx.doi.org/10.1002/prop.201300035}{{\em
  Fortsch. Phys.} {\bfseries 62} (2014) 155--206},
  \href{http://arxiv.org/abs/1310.1325}{{\ttfamily arXiv:1310.1325 [hep-th]}}.

\bibitem{Hohenegger:2015cba}
S.~Hohenegger, A.~Iqbal, and S.-J. Rey, ``{M-strings, monopole strings, and
  modular forms},'' \href{http://dx.doi.org/10.1103/PhysRevD.92.066005}{{\em
  Phys. Rev. D} {\bfseries 92} no.~6, (2015) 066005},
  \href{http://arxiv.org/abs/1503.06983}{{\ttfamily arXiv:1503.06983
  [hep-th]}}.

\bibitem{Hohenegger:2015btj}
S.~Hohenegger, A.~Iqbal, and S.-J. Rey, ``{Instanton-monopole correspondence
  from M-branes on $\mathbb S^1$ and little string theory},''
  \href{http://dx.doi.org/10.1103/PhysRevD.93.066016}{{\em Phys. Rev. D}
  {\bfseries 93} no.~6, (2016) 066016},
  \href{http://arxiv.org/abs/1511.02787}{{\ttfamily arXiv:1511.02787
  [hep-th]}}.

\bibitem{Katz:2002gh}
S.~H. Katz and E.~Sharpe, ``{D-branes, open string vertex operators, and Ext
  groups},'' \href{http://dx.doi.org/10.4310/ATMP.2002.v6.n6.a1}{{\em Adv.
  Theor. Math. Phys.} {\bfseries 6} (2003) 979--1030},
  \href{http://arxiv.org/abs/hep-th/0208104}{{\ttfamily arXiv:hep-th/0208104}}.

\bibitem{Gomis:2008vc}
J.~Gomis, D.~Rodriguez-Gomez, M.~Van~Raamsdonk, and H.~Verlinde, ``{A Massive
  Study of M2-brane Proposals},''
  \href{http://dx.doi.org/10.1088/1126-6708/2008/09/113}{{\em JHEP} {\bfseries
  09} (2008) 113}, \href{http://arxiv.org/abs/0807.1074}{{\ttfamily
  arXiv:0807.1074 [hep-th]}}.

\bibitem{Gaiotto:2008ak}
D.~Gaiotto and E.~Witten, ``{S-Duality of Boundary Conditions In N=4 Super
  Yang-Mills Theory},''
  \href{http://dx.doi.org/10.4310/ATMP.2009.v13.n3.a5}{{\em Adv. Theor. Math.
  Phys.} {\bfseries 13} no.~3, (2009) 721--896},
  \href{http://arxiv.org/abs/0807.3720}{{\ttfamily arXiv:0807.3720 [hep-th]}}.

\bibitem{Gopakumar:1998ii}
R.~Gopakumar and C.~Vafa, ``{M theory and topological strings. 1.},''
  \href{http://arxiv.org/abs/hep-th/9809187}{{\ttfamily arXiv:hep-th/9809187}}.

\bibitem{Gopakumar:1998jq}
R.~Gopakumar and C.~Vafa, ``{M theory and topological strings. 2.},''
  \href{http://arxiv.org/abs/hep-th/9812127}{{\ttfamily arXiv:hep-th/9812127}}.

\bibitem{Hollowood:2003cv}
T.~J. Hollowood, A.~Iqbal, and C.~Vafa, ``{Matrix models, geometric engineering
  and elliptic genera},''
  \href{http://dx.doi.org/10.1088/1126-6708/2008/03/069}{{\em JHEP} {\bfseries
  03} (2008) 069}, \href{http://arxiv.org/abs/hep-th/0310272}{{\ttfamily
  arXiv:hep-th/0310272}}.

\bibitem{Sharpe:2003dr}
E.~Sharpe, ``{Lectures on D-branes and sheaves},''
\newblock 7, 2003.
\newblock \href{http://arxiv.org/abs/hep-th/0307245}{{\ttfamily
  arXiv:hep-th/0307245}}.

\bibitem{Hohenegger:2016eqy}
S.~Hohenegger, A.~Iqbal, and S.-J. Rey, ``{Self-Duality and Self-Similarity of
  Little String Orbifolds},''
  \href{http://dx.doi.org/10.1103/PhysRevD.94.046006}{{\em Phys. Rev. D}
  {\bfseries 94} no.~4, (2016) 046006},
  \href{http://arxiv.org/abs/1605.02591}{{\ttfamily arXiv:1605.02591
  [hep-th]}}.

\bibitem{DelZotto:2014hpa}
M.~Del~Zotto, J.~J. Heckman, A.~Tomasiello, and C.~Vafa, ``{6d Conformal
  Matter},'' \href{http://dx.doi.org/10.1007/JHEP02(2015)054}{{\em JHEP}
  {\bfseries 02} (2015) 054}, \href{http://arxiv.org/abs/1407.6359}{{\ttfamily
  arXiv:1407.6359 [hep-th]}}.

\bibitem{Morrison:2012np}
D.~R. Morrison and W.~Taylor, ``{Classifying bases for 6D F-theory models},''
  \href{http://dx.doi.org/10.2478/s11534-012-0065-4}{{\em Central Eur. J.
  Phys.} {\bfseries 10} (2012) 1072--1088},
  \href{http://arxiv.org/abs/1201.1943}{{\ttfamily arXiv:1201.1943 [hep-th]}}.

\bibitem{Okazaki:2015fiq}
T.~Okazaki and D.~J. Smith, ``{Topological M-strings and supergroup
  Wess-Zumino-Witten models},''
  \href{http://dx.doi.org/10.1103/PhysRevD.94.065016}{{\em Phys. Rev. D}
  {\bfseries 94} no.~6, (2016) 065016},
  \href{http://arxiv.org/abs/1512.06646}{{\ttfamily arXiv:1512.06646
  [hep-th]}}.

\bibitem{Okazaki:2016pne}
T.~Okazaki and D.~J. Smith, ``{Mock modular index of M2-M5 brane systems},''
  \href{http://dx.doi.org/10.1103/PhysRevD.96.026017}{{\em Phys. Rev. D}
  {\bfseries 96} no.~2, (2017) 026017},
  \href{http://arxiv.org/abs/1612.07565}{{\ttfamily arXiv:1612.07565
  [hep-th]}}.

\bibitem{Witten:1986bf}
E.~Witten, ``{Elliptic Genera and Quantum Field Theory},''
  \href{http://dx.doi.org/10.1007/BF01208956}{{\em Commun. Math. Phys.}
  {\bfseries 109} (1987) 525}.

\bibitem{Witten:1993jg}
E.~Witten, ``{On the Landau-Ginzburg description of N=2 minimal models},''
  \href{http://dx.doi.org/10.1142/S0217751X9400193X}{{\em Int. J. Mod. Phys. A}
  {\bfseries 9} (1994) 4783--4800},
  \href{http://arxiv.org/abs/hep-th/9304026}{{\ttfamily arXiv:hep-th/9304026}}.

\bibitem{Berglund:1993fj}
P.~Berglund and M.~Henningson, ``{Landau-Ginzburg orbifolds, mirror symmetry
  and the elliptic genus},''
  \href{http://dx.doi.org/10.1016/0550-3213(94)00389-V}{{\em Nucl. Phys. B}
  {\bfseries 433} (1995) 311--332},
  \href{http://arxiv.org/abs/hep-th/9401029}{{\ttfamily arXiv:hep-th/9401029}}.

\bibitem{Henningson:1993nr}
M.~Henningson, ``{N=2 gauged WZW models and the elliptic genus},''
  \href{http://dx.doi.org/10.1016/0550-3213(94)90614-9}{{\em Nucl. Phys. B}
  {\bfseries 413} (1994) 73--83},
  \href{http://arxiv.org/abs/hep-th/9307040}{{\ttfamily arXiv:hep-th/9307040}}.

\bibitem{Nemeschansky:1994ac}
D.~Nemeschansky and N.~P. Warner, ``{The Refined elliptic genus and Coulomb gas
  formulations of N=2 superconformal coset models},''
  \href{http://dx.doi.org/10.1016/0550-3213(95)00059-2}{{\em Nucl. Phys. B}
  {\bfseries 442} (1995) 623--654},
  \href{http://arxiv.org/abs/hep-th/9412187}{{\ttfamily arXiv:hep-th/9412187}}.

\bibitem{Hosomichi:2014rqa}
K.~Hosomichi and S.~Lee, ``{Self-dual Strings and 2D SYM},''
  \href{http://dx.doi.org/10.1007/JHEP01(2015)076}{{\em JHEP} {\bfseries 01}
  (2015) 076}, \href{http://arxiv.org/abs/1406.1802}{{\ttfamily arXiv:1406.1802
  [hep-th]}}.

\bibitem{Gadde:2013dda}
A.~Gadde and S.~Gukov, ``{2d Index and Surface operators},''
  \href{http://dx.doi.org/10.1007/JHEP03(2014)080}{{\em JHEP} {\bfseries 03}
  (2014) 080}, \href{http://arxiv.org/abs/1305.0266}{{\ttfamily arXiv:1305.0266
  [hep-th]}}.

\bibitem{Gadde:2015tra}
A.~Gadde, B.~Haghighat, J.~Kim, S.~Kim, G.~Lockhart, and C.~Vafa, ``{6d String
  Chains},'' \href{http://dx.doi.org/10.1007/JHEP02(2018)143}{{\em JHEP}
  {\bfseries 02} (2018) 143}, \href{http://arxiv.org/abs/1504.04614}{{\ttfamily
  arXiv:1504.04614 [hep-th]}}.

\bibitem{Benini:2018hjy}
F.~Benini, G.~Bonelli, M.~Poggi, and A.~Tanzini, ``{Elliptic non-Abelian
  Donaldson-Thomas invariants of $\mathbb{C}^3$},''
  \href{http://arxiv.org/abs/1807.08482}{{\ttfamily arXiv:1807.08482
  [hep-th]}}.

\bibitem{Nakajima:2003pg}
H.~Nakajima and K.~Yoshioka, ``{Instanton counting on blowup. 1.},''
  \href{http://dx.doi.org/10.1007/s00222-005-0444-1}{{\em Invent. Math.}
  {\bfseries 162} (2005) 313--355},
  \href{http://arxiv.org/abs/math/0306198}{{\ttfamily arXiv:math/0306198}}.

\end{thebibliography}\endgroup

\end{document}